\documentclass[twocolumn]{autart} 

\newtheorem{mydef}{\bf Definition}
\newtheorem{mythm}{\bf Theorem}
\newtheorem{myprob}{\bf Problem}
\newtheorem{mylem}{\bf Lemma}

\newtheorem{mypro}{\bf Proposition}

\newtheorem{remark}{Remark}
\newtheorem{assumption}{\bf Assumption}

\usepackage{amsfonts}
\DeclareMathSymbol{\shortminus}{\mathbin}{AMSa}{"39}

\usepackage{flushend}
\usepackage{multirow}
\usepackage{makecell}
\usepackage{booktabs}
\usepackage{array}

\usepackage{graphicx}
\graphicspath{{./Pics/}}
\DeclareGraphicsExtensions{.pdf}
\usepackage{float}
\usepackage{amsmath}

\usepackage{epsfig}
\usepackage{graphicx}
\usepackage{arydshln}
\usepackage{algpseudocode}
\usepackage{verbatim}
\usepackage{subfigure}
\usepackage{enumerate}
\usepackage{rotating}\usepackage{enumitem}
\usepackage{color}
\usepackage{caption} 
\usepackage{mathrsfs}
\usepackage{amssymb}
\usepackage{mathtools}
\usepackage[linesnumbered,ruled,vlined]{algorithm2e}
\usepackage{cite}
\usepackage{float}
\usepackage{soul}
\usepackage{tikz}
\usepackage{hyperref}
\hypersetup{hidelinks} 
\usepackage{soul}
\usepackage{booktabs}
\usepackage{multirow}
\usepackage{makecell}
\usepackage{array}
\usetikzlibrary{arrows.meta}
\usetikzlibrary{patterns}
\usepackage{tcolorbox}
\usepackage{fancyhdr}
\usepackage{flushend}
\allowdisplaybreaks[4]

\definecolor{myblue}{RGB}{135,206,235}
\definecolor{myyellow}{RGB}{255,255,0}
\definecolor{mygreen}{RGB}{0,255,0}

\floatname{algorithm}{Procedure}

\begin{document}
\begin{frontmatter}
\title{Control Synthesis for Multiple Reach-Avoid Tasks\\ via Hamilton-Jacobi Reachability Analysis\vspace{-6pt}}

\thanks[footnoteinfo]{This work was supported by  the National Natural Science Foundation of China (62061136004, 62173226, 61833012).}
		\author[SJTU1]{Yu Chen}\ead{yuchen26@sjtu.edu.cn},  
		\author[SJTU1]{Shaoyuan Li}\ead{syli@sjtu.edu.cn},
		\author[SJTU1]{Xiang Yin}\ead{yinxiang@sjtu.edu.cn} 
		
		\address[SJTU1]{School of Automation and Intelligent Sensing, Shanghai Jiao Tong University, Shanghai 200240, China.} 
\begin{abstract}
We investigate the control synthesis problem for continuous-time time-varying nonlinear systems with disturbance under a class of \emph{multiple reach-avoid} (MRA) tasks. Specifically, the MRA task requires the system to reach a series of target regions in a specified order while satisfying state constraints between each pair of target arrivals. This problem is more challenging than standard reach-avoid tasks, as it requires considering the feasibility of future reach-avoid tasks during the planning process.  
To solve this problem, we define a series of value functions by solving a cascade of time-varying reach-avoid problems characterized by  Hamilton-Jacobi variational inequalities. 
We prove that the super-level set of the final value function computed is exactly the feasible set of the MRA task. Additionally, we demonstrate that the control law can be effectively synthesized by ensuring the non-negativeness of the value functions over time.
We also show that the Linear temporal logic task control synthesis problems can be converted to a collection of MRA task control synthesis problems by properly defining each target and state constraint set of MRA tasks.
The effectiveness of the proposed approach is illustrated through four case studies on robot planning problems under time-varying nonlinear systems with disturbance.
\end{abstract}
\end{frontmatter}

\section{Introduction} 
\subsection{Motivations} 
Formal controller synthesis is a fundamental challenge in autonomous systems such as autonomous driving \cite{mehdipour2023formal}, marine surface vessels \cite{dietrich2025symbolic}, and industrial manufacturing systems \cite{campos2018formal}. The objective is to algorithmically design a controller that can formally guarantee the satisfaction of a given specification, with mathematically rigorous proofs. Over the years, formal controller synthesis with provable guarantees has been extensively studied for various system classes and requirement types, as highlighted in recent surveys \cite{belta2019formal, pola2019control, yin2024formal}. As autonomous systems become increasingly complex, ensuring their safe and efficient operation requires more advanced synthesis techniques, ranging from improvements in scalability to the incorporation of additional functionality.

Among formal specifications, \emph{reachability} is one of the most fundamental tasks. It requires the system to reach a target state either eventually or within a predefined time horizon. In many applications, additional constraints are imposed, such as avoiding obstacles or remaining within the target region once it has been reached. These tasks are commonly referred to as reach-avoid~\cite{margellos2011hamilton, fisac2015reach, zhou2018efficient, xue2024reach} or reach-avoid-stay problems~\cite{meng2022smooth, meng2024stochastic, das2024prescribed}. Reach-avoid tasks are not only important in their own right but also serve as fundamental building blocks for more complex specifications. For example, in linear temporal logic (LTL) specifications, a system must sequentially reach specific labeled regions according to automata states in order to satisfy the desired temporal-spatial behavior \cite{zhong2024formal}.

In formal controller synthesis for reachability-based specifications, the key challenge lies in analyzing reachability based on system dynamics. 
This problem can be addressed using Hamilton-Jacobi Reachability (HJR) analysis, which formulates it as a Hamilton-Jacobi partial differential equation (PDE)~\cite{bansal2017hamilton} and represents the region of interest as the level set of the PDE solution. The HJR method provides a theoretical foundation for synthesizing controllers for reachability~\cite{mitchell2005time} and reach-avoid~\cite{fisac2015reach} tasks in dynamic systems under disturbances. 
In recent years, it has been applied to a wide range of complex reach-avoid problems, including multiplayer reach-avoid games~\cite{huang2011differential} and multi-vehicle path planning~\cite{chen2018robust}. 

\subsection{Our Results}
In this paper, we address the control synthesis problem for time-varying nonlinear systems under a new class of tasks called the \emph{multiple reach-avoid} (MRA) task. 
An MRA task requires the system to reach a series of (time-varying) targets in a predefined sequence while satisfying state constraints between each pair of consecutive arrivals. 
This problem is more challenging than standard reach-avoid tasks, as it necessitates ensuring the feasibility of future reach-avoid sub-tasks during the planning process. 
Such task can thus be regarded as a foundational problem for solving more complex temporal tasks.

Our key results and contributions are summarized as follows:\vspace{-3pt}
\begin{itemize}[leftmargin=*,topsep=0pt, itemsep=5pt]
\item
First, we prove that the feasible set of the MRA task can be exactly characterized as the super-level set of a specific function. This function is computed by solving a sequence of time-varying reach-avoid Hamilton-Jacobi Reachability (HJR) problems, where the feasible set of future sub-tasks is treated as a dynamic target. Our method provides a new perspective on how to extend simple reach-avoid tasks to more complex temporal tasks.
\item
We then propose an efficient online algorithm for selecting control inputs to achieve MRA tasks. This is done by ensuring that the value functions remain non-negative over time. The algorithm can be viewed as a filter for MRA task satisfaction and is compatible with controllers designed for other performance objectives.
\item
Additionally, we discuss how the proposed MRA framework is related to linear temporal logic  specifications through its automata representation. 
Specifically, we demonstrate that by properly defining each target and state constraint set, an MRA task can be used as a sound approach for synthesizing controllers for LTL tasks.
This further highlights that MRA tasks serve as a fundamental building block for addressing more complex temporal logic tasks.
\item
Finally, we provide a comprehensive set of case studies and simulations, ranging from mobile robot task planning to spacecraft rendezvous.
These demonstrate the effectiveness of our method in different types of nonlinear systems with potential disturbances.
\end{itemize}

\subsection{Related Works}
Our work is related to solving control synthesis problems under complex temporal requirements using the HJR method; see, e.g., \cite{chen2018signal,gao2021temporal,yu2024continuous,verhagen2024robust,jiang2024guaranteed}. For instance, \cite{chen2018signal} computes the feasible sets of signal temporal logic tasks by recursively handling each temporal operator, while \cite{gao2021temporal,yu2024continuous,verhagen2024robust,jiang2024guaranteed} introduce a temporal logic tree structure to heuristically guide the feasible set computation for linear temporal logic tasks. These approaches rely on heuristic algorithms to account for temporal task dependencies, yielding only conservative approximations of the feasible set. In contrast, our work provides an exact functional representation of the proposed MRA task by treating the feasible set of future sub-tasks as a time-varying dynamic target. This approach offers new insights into precisely characterizing complex temporal dependencies through the HJR method.

Our method presents a sound approach for LTL control synthesis of nonlinear systems. Existing works on this topic have developed various techniques, such as abstraction-based approaches~\cite{liu2016finite,reissig2016feedback}, control barrier functions~\cite{srinivasan2020control}, and optimization-based approaches~\cite{wolff2014optimization}. Our method can be applied to time-varying tasks under possible disturbances, while existing methods, with the exception of abstraction-based methods, can only handle deterministic systems. Moreover, only systems with favorable properties (e.g., incremental stability) admit guaranteed finite abstractions, whereas our method can be applied to general control-affine systems without further assumptions. Furthermore, our method computes the feasible control input set for task satisfaction, which is compatible with other reference controllers.

\subsection{Organizations}
The rest of the paper is organized as follows. 
We present some preliminaries in Section \ref{sec:prelinimary} and 
formulate the problem in Section \ref{sec:problemformulation}. 
Section \ref{sec:functioncomputation} presents how to compute the feasible set for an MRA task  by solving a series of HJ variational inequalities. 
These function are then used to solve the controller synthesis problem in Section~\ref{sec:controlsynthesis}. In Section~\ref{sec:discussall relation}, we discuss how our approach can be applied to the LTL controller synthesis problem.
Finally, we illustrate the proposed method by four case studies in Section~\ref{sec:simulation} and conclude the paper in Section~\ref{sec:con}.

A preliminary and partial version of this paper was presented in \cite{cdc2025chen}. Compared with the conference version, the present journal version has the following main differences. First, this paper considers a setting of time-varying task under disturbances, whereas the conference paper only considers a static and deterministic setting. Second, we demonstrate how our approach can be applied to LTL control synthesis by suitably defining an MRA task. Furthermore, this work provides extensive case studies and simulations to illustrate the effectiveness of the proposed method. Additionally, we present the complete proof, which were either not included or only sketched in the conference version.
\section{Preliminary}\label{sec:prelinimary}
\textbf{Notations: }
Let $A$ be a finite set of symbols. 
We denote by $A^{*}$  and $A^\omega$ the sets of all finite and infinite sequences over $A$, respectively, and $\epsilon \in A^*$ is the empty sequence. The power set of $A$ is denoted by $2^A$.
%We say $\rho \in A^*$ is a \emph{simple sequence} if all symbols of $\rho$ are different. 
We denote by $\mathbb{R}$, $\mathbb{R}_{\geq 0}$ and $\mathbb{R}^n$ the set of all real numbers, non-negative real numbers and $n$-dimensional real vectors, respectively.  
\subsection{System and Trajectories} 
We consider a nonlinear system described by
\begin{equation} \label{eq:systemdynamic}
    %\frac{\mathrm{d}x}{\mathrm{d}t}:=
    \dot{x}(t)= f(x,t)+g(x,t)u+p(x,t)d=\Bar{f}(x,u,d,t),
\end{equation}
where 
$x \in \mathbb{R}^n$ is the system state, 
$u \in \mathcal{U} \subseteq \mathbb{R}^m$ is the control input with compact input space $\mathcal{U}$, 
$d \in \mathcal{D}\subseteq \mathbb{R}^l$ is the disturbance with compact disturbance space $\mathcal{D}$, 
and
$f: \mathbb{R}^n \times \mathbb{R}_{\geq 0} \to \mathbb{R}^n$, $g: \mathbb{R}^n\times \mathbb{R}_{\geq 0} \to \mathbb{R}^{n \times m}$, $p:\mathbb{R}^n\times\mathbb{R}_{\geq 0} \to \mathbb{R}^{n\times l}$ are bounded functions with uniformly continuous in $t$ and Lipschitz continuous in $x$. 

The set of admissible control functions over time interval $[t_0, t_1]$ with $0\leq t_0\leq t_1$ is defined as  
\begin{equation}  
    \mathbb{U}_{[t_0,t_1]} := \{ \mathbf{u}: [t_0, t_1] \to \mathcal{U} \mid \mathbf{u}(\cdot) \text{ is measurable} \}.   
\end{equation}
Similarly,  the set of admissible disturbance functions over time interval $[t_0,t_1]$ is defined as  
\begin{equation}  
    \mathbb{D}_{[t_0,t_1]} := \{ \mathbf{d}: [t_0, t_1] \to \mathcal{D} \mid \mathbf{d}(\cdot) \text{ is measurable} \}.   
\end{equation}
For an initial state $x \in \mathbb{R}^n$ and time instant $t$, under   control function $\mathbf{u} \in \mathbb{U}_{[t,s]}$ and  disturbance function $\mathbf{d} \in \mathbb{D}_{[t,s]}$, the  evolution of the systems is determined by the unique continuous trajectory $\xi_{x,t}^{\mathbf{u},\mathbf{d}}: [t,s] \to \mathbb{R}^n$ such that $\xi_{x,t}^{\mathbf{u},\mathbf{d}}(t) = x$ and  
\begin{equation}  
    \dot{\xi}_{x,t}^{\mathbf{u},\mathbf{d}}(\tau) = 
   \Bar{f}(\xi_{x,t}^{\mathbf{u},\mathbf{d}}(\tau),\mathbf{u}(\tau),\mathbf{d}(\tau),\tau), \ \text{a.e. } \tau \in [t,s],   
\end{equation} 
where a.e. (almost everywhere) means the differential equation holds except on a set of Lebesgue measure zero.  
During the online execution, a feedback control policy is used. 
Let $\mathbf{c}:\mathbb{R}^n\times [0,T]\to \mathcal{U}$ be a state-feedback control function which is uniformly continuous in $t$ and Lipschitz continuous in $x$. Then we can similarly denote by $\xi_{x,t}^{\mathbf{c},\mathbf{d}}: [t,s]\to \mathbb{R}^n$ the system trajectory when $\mathbf{u}$ is replaced by $\mathbf{c}.$

In this work, we allow the the adversarial environment to take non-anticipative strategies~\cite{bansal2017hamilton,chen2018hamilton} defined as
\begin{align}
&\Gamma_{[t_0,t_1]}= \\
&\left\{
 \gamma\!: \! \mathbb{U}_{[t_0,t_1]}\!\!\to\!\!\mathbb{D}_{[t_0,t_1]} \mid\!\!\!
\begin{array}{l l}
     &  \forall \mathbf{u},\hat{\mathbf{u}} \in \mathbb{U}_{[t_0,t_1]}, \forall s \in [t_0,t_1]\\
     &  [ \mathbf{u}(\tau)\!=\!\hat{\mathbf{u}}(\tau) \text{ a.e. } \tau \!\in\! [t_0,s] ]\Rightarrow \\
     &[\gamma[\mathbf{u}](\tau)\!=\!\gamma[\hat{\mathbf{u}}](\tau) \text{ a.e. } \tau \!\in \![t_0,s] ]
\end{array} 
\right\}.\nonumber
\end{align} 
Intuitively, non-anticipative strategies allow the environment to make decisions about $\mathbf{d}(s)$ with full knowledge of $\mathbf{u}(\tau)$ for $\tau \in [t_0,s]$. In this setting, the environment has access to both the state feedback and the current control input, while the control function can only use the state feedback information available up to the current time. 
For the sake of simplicity, 
we denote by $\xi^{\mathbf{u},\gamma}_{x,t}$ the trajectory under control function $\mathbf{u}$ and strategy $\gamma$, i.e., 
$\xi^{\mathbf{u},\gamma(\mathbf{u})}_{x,t}$.

\subsection{Time-Varying Reachability}
In a reach-avoid task, the system needs to reach a target region while remaining within a safe region throughout its trajectory.  
Formally, a (time-varying) reach-avoid task is defined as a tuple $(\mathcal{T}, \mathcal{G})$, where $\mathcal{T}, \mathcal{G} \subseteq \mathbb{R}^n \times [t_0, t_1]$ are time-augmented sets representing the target region and the safe region, respectively.  
For each $\star \in \{\mathcal{T}, \mathcal{G}\}$ and time instant $t \in [t_0, t_1]$,  
we denote the state set at time $t$ by  
$\star(t) = \{ x \in \mathbb{R}^n \mid (x, t) \in \star \}$.
Given $t_0 \leq t \leq t_1$, the \emph{feasible set} of the reach-avoid task $(\mathcal{T}, \mathcal{G})$ is defined as  
\begin{align}\label{eq:HJRvalueequvialent}   
&\mathsf{RA}(t, t_1, \mathcal{T}, \mathcal{G}) =\\
&\left\{\!\! x\! \in \! \mathbb{R}^n \!\!\mid \!\!\! 
\begin{array}{cc} 
     (\forall \gamma \!\in\! \Gamma_{[t,t_1]} )
     (\exists \mathbf{u} \!\in \!\mathbb{U}_{[t, t_1]})(\exists s \!\in\! [t, t_1])\\
     \bigl[\xi^{\mathbf{u},\gamma}_{x,t}(s)  \!\in\! \mathcal{T}(s)\bigr]\!\wedge \! 
     [\forall s' \!\in\! [t, s]\!:\! \xi^{\mathbf{u},\gamma}_{x,t}(s') \!\in\! \mathcal{G}(s')]
\end{array}
\!\! \right\}.  \nonumber
\end{align}
That is, a state is in the feasible set of reach-avoid task iff initial from state $x$ at time $t$, regardless of the non-anticipative strategies of environment, we can find a control function such that, the system trajectory can reach target at some time instant $s$ before $t_1$ and always stay in safe region before $s$.
For each $\star \in \{ \mathcal{T},\mathcal{G}\}$, 
we assume that  there is Lipschitz continuous function 
$h_{\star}: \mathbb{R}^n \times [t_0,t_1] \to \mathbb{R}$ 
such that 
$h_{\star}(x,t) \geq 0$ iff $x \in \star(t)$.
Then we define a value function by:  $\forall x\in \mathbb{R}^n, t \in [t_0,t_1]$, we have
\begin{align} \label{eq:defofHJRRA}
 &h_{\mathsf{RA}}(x,t,h_{\mathcal{T}},h_{\mathcal{G}})= 
 \mathop{\inf}_{\gamma \in \Gamma_{[t,t_1]}}\mathop{\sup}_{\mathbf{u}\in \mathbb{U}_{[t,t_1]} } \\
 &  \mathop{\max}_{s \in [t,t_1] }  \min 
\left\{ 
 h_{\mathcal{T}}(\xi_{x,t}^{\mathbf{u},\gamma}(s), s),   
\min_{s' \in [t,s]} h_{\mathcal{G}}(\xi_{x,t}^{\mathbf{u},\gamma}(s'), s')  
\right\} .\nonumber
\end{align}
According to \cite{fisac2015reach}, we know that   $x \in \mathsf{RA}(t,t_1,\mathcal{T},\mathcal{G})$ iff $h_{\mathsf{RA}}(x,t,h_{\mathcal{T}},h_{\mathcal{G}}) \geq 0$.
Moreover, the value function $h_{\mathsf{RA}}$ is the viscosity solution of following Hamilton-Jacobi variational inequality (VI):
\begin{align} \label{eq:HJR}
&\min\left\{\!\!\!\!\!
\begin{array}{cc}
     &  \max 
     \{  
     \frac{\partial h_{\mathsf{RA}}(x,t)}{\partial t}  +  \mathsf{Ham}(x,t),    h_{\mathcal{T}}(x,t)-h_{\mathsf{RA}}(x,t) 
     \!\},\\
     & h_{\mathcal{G}}(x,t) -h_{\mathsf{RA}}(x,t)
\end{array}\!\!
\right\}\nonumber\\
&=0,
\end{align}
where 
$h_{\mathsf{RA}}(x,t_1)=\min\{ h_{\mathcal{T}}(x,t_1), h_{\mathcal{G}}(x,t_1) \}$ is the boundary condition and 
\begin{align}
&    \mathsf{Ham}(x,t) =\\
& \max_{u \in \mathcal{U} } \min_{d \in \mathcal{D}} \frac{\partial h_{\mathsf{RA}}(x,t)}{\partial x} (f(x,t)+g(x,t)u+p(x,t)d).\nonumber
\end{align} 
The reader is referred to~\cite{fisac2015reach} for more details on solving time-varying reachability problem by HJR method.

\section{Problem Formulation}\label{sec:problemformulation}
In this work, we introduce a new type of reach-avoid task called the \emph{multiple reach-avoid} (MRA) task.  
The objective of the system is  to visit \emph{a sequence of} target regions in a prescribed order while satisfying state constraints between consecutive target arrivals.  
Formally, let
\[
\mathbb{T}=(\mathcal{T}_1,\mathcal{T}_2,\dots,\mathcal{T}_N)\text{ and }\mathbb{G}=(\mathcal{G}_1,\mathcal{G}_2,\dots, \mathcal{G}_N)
\]
denote sequences of $N$ target regions and $N$ safe regions, respectively, 
where for each $i=1,\dots,N$, the sets satisfy $\mathcal{T}_i, \mathcal{G}_i \subseteq \mathbb{R}^n\times [0,T]$. 
Let $t_0, t_1 \in [0, T]$ represent the start time and end time of the entire task.  
The \emph{multiple reach-avoid task} (MRA) is then defined by the 4-tuple:  
\[
\Phi = (t_0, t_1, \mathbb{T}, \mathbb{G}).
\]  
For simplicity, we denote an MRA task by $\Phi^{[t_0,t_1]}$ when the target and safe regions are clear from the context. 
Given an initial state $x_0 \in \mathbb{R}^n$, a control function $\mathbf{u} \in \mathbb{U}_{[t_0,t_1]}$, and a disturbance function $\mathbf{d} \in \mathbb{D}_{[t_0,t_1]}$, the generated trajectory $\xi_{x_0,t_0}^{\mathbf{u},\mathbf{d}}$ is said to satisfy the MRA task $\Phi^{[t_0,t_1]}$, denoted by $\xi_{x_0,t_0}^{\mathbf{u},\mathbf{d}} \models \Phi^{[t_0,t_1]}$, if there exists a sequence of time instants  
\[
t_0 = \tau_0 \leq \tau_1 \leq \tau_2 \leq \cdots \leq \tau_N \leq t_1  
\]  
such that $\forall i=1,\dots,N$, we have
\begin{equation}\label{eq:taskdes}  
 \bigg[\xi_{x_0,t_0}^{\mathbf{u},\mathbf{d}}(\tau_i) \!\in\! \mathcal{T}_i(\tau_i)\bigg]  
\wedge  
\bigg[\forall \tau \!\in\! [\tau_{i-1}, \tau_i]: \xi_{x_0,t_0}^{\mathbf{u},\mathbf{d}}(\tau) \!\in\! \mathcal{G}_i(\tau) \bigg].  
\end{equation}  
The feasible set of task $\Phi^{[t_0,t_1]} $ is defined as
\begin{align}\label{eq:MRAfeasible}
& \mathsf{MRA}(t_0, t_1, \mathbb{T}, \mathbb{G}) = \\
&\Big\{ x  \in \mathbb{R}^n \mid \forall \gamma \in \Gamma_{[t_0,t_1]},  
  \exists \mathbf{u} \in \mathbb{U}_{[t_0,t_1]}, \xi^{\mathbf{u},\gamma}_{x,t_0} \models \Phi^{[t_0,t_1]} \Big\}. \nonumber
\end{align}
We now state the MRA task control synthesis problem.
\begin{myprob}[MRA Control Synthesis] \label{problem}\upshape
 Given the nonlinear system \eqref{eq:systemdynamic}, an initial state \( x_0 \in \mathbb{R}^n \), and an MRA task \( \Phi = (0, T, \mathbb{T}, \mathbb{G}), \)\vspace{-6pt} 
\begin{itemize}
    \item[(1)] 
    Decide whether \( x_0 \in \mathsf{MRA}(0, T, \mathbb{T}, \mathbb{G}) \). \vspace{3pt} 
    \item[(2)] 
    If so, find a state-feedback control function $\mathbf{c}:\mathbb{R}^n\times[0,T]\to \mathcal{U}$ such that $\forall \mathbf{d} \in \mathbb{D}_{[0,T]}, \xi_{x_0,0}^{\mathbf{c},\mathbf{d}} \models \Phi$.
\end{itemize} 
\end{myprob} 

\begin{remark}\upshape
According to the definition of feasible sets, it appears that the decision space of the controller is $\mathbb{U}_{[0,T]}$. However, under a non-anticipative strategy, both the controller and the disturbance have knowledge of each other's decisions up to the current time step. That is, both players are aware of the current system state. This implies that, during the controller design phase, instead of seeking a function $\mathbf{u} \in \mathbb{U}_{[0,T]}$, we should find a state-feedback control function $\mathbf{c}: \mathbb{R}^n \times [0,T] \to \mathcal{U}$ to reactively handle the bounded disturbance. Therefore, most existing works~\cite{bansal2017hamilton, chen2018hamilton, 9683085} based on HJR for safe control use this dynamic-game-based formulation to describe the feasible set of tasks, while still focusing on finding a state-feedback control function during the control synthesis stage. Our work also adopts the conventional formulation, consistent with existing literature.
\end{remark} 
\section{Value Function Computation}\label{sec:functioncomputation}
In this section, we adopt the HJR method to compute the value function, whose super-level set represents the feasible set of the MRA task $\Phi^{[0,T]}$. 
Particularly, if the initial state lies within this feasible set, then the value function will later be used to ensure task satisfaction (as discussed in the next section). 
Note that the standard HJR method computes the feasible set only for a single reach-avoid task. 
We show here this approach can be extended to the MRA task by treating the feasible set of future tasks as a \emph{dynamic target}. 
First, we assume that the target and safe regions can be expressed in terms of value functions, as formalized below.  
\begin{assumption} \label{assum:global}\upshape
Let  
$
\mathbb{T} = (\mathcal{T}_1, \mathcal{T}_2, \dots, \mathcal{T}_N)  
$  
and  
$
\mathbb{G} = (\mathcal{G}_1, \mathcal{G}_2, \dots, \mathcal{G}_N)  
$ 
be sequences of $N$ target regions and  $N$ safe regions, respectively, where $\mathcal{T}_i, \mathcal{G}_i \subseteq \mathbb{R}^n\times [0,T]$.
For each $i=1,2,\dots,N$, we assume there exist Lipschitz continuous functions  
\begin{equation}
    h_{\mathcal{T}_i}, h_{\mathcal{G}_i} \colon \mathbb{R}^n \times [0,T] \to \mathbb{R}  
\end{equation} 
that characterize the target region $\mathcal{T}_i$ and safe region $\mathcal{G}_i$, respectively, such that for each $\star \in \{\mathcal{T}, \mathcal{G}\}$, we have   
\begin{equation}  
 (x,t) \in \star_i(t) \Leftrightarrow h_{\star_i}(x,t) \geq 0 .  
\end{equation} 
\end{assumption}
Our approach for computing the overall value function for the MRA task consists of the following steps: \vspace{-3pt} 
\begin{enumerate}  
    \item 
    \textbf{Initial Feasible Set Computation.}   
    First, we compute the feasible set for a single-target reach-avoid task with target function \( h_{\mathcal{T}_N} \) and safe function \( h_{\mathcal{G}_N} \), leveraging existing results from \cite{fisac2015reach} (formally stated in Lemma~\ref{lem:firstcase}).  \medskip
    \item 
    \textbf{Recursive Value Function Construction.}   
    Next, we compute a new value function for a new single-target reach-avoid task by treating:  
    (i) the value function from the previous step and \( h_{\mathcal{T}_{N-1}} \) as a combined \emph{time-varying} target function, and  
    (ii) \( h_{\mathcal{G}_{N-1}} \) as the safe function.  
    This value function represents the feasible set for first reaching \( \mathcal{T}_{N-1} \) and then \( \mathcal{T}_N \), while remaining in \( \mathcal{G}_{N-1} \) and \( \mathcal{G}_N \) before arriving at \( \mathcal{T}_{N-1} \) and \( \mathcal{T}_N \), respectively. A formal proof is provided in Proposition~\ref{prop:secondcase}.  \medskip
    \item 
    \textbf{Iterative Extension to Full MRA Task.}   
    We repeat the above step recursively, treating the value function computed at each iteration as a time-varying target for the next. The remaining target \( h_{\mathcal{T}_{i}} \) and safe function \( h_{\mathcal{G}_{i}} \) (not yet considered) are incorporated as the new target and safe regions, respectively.  
    Upon including all targets, the super-level set of the final value function yields the feasible set of the MRA task, as proven in Theorem~\ref{thm:secondcombine}.  
\end{enumerate}  
To formally establish our results, we define 
\[
\mathbb{T}_i = (\mathcal{T}_{N-i+1}, \dots, \mathcal{T}_N)
\]
as the sequence of the last $i$ target sets, i.e.,  
 $\mathbb{T}_1$ means that one only needs to achieve the last task $\mathcal{T}_N$. 
Similarly, we define 
$\mathbb{G}_i = (\mathcal{G}_{N-i+1}, \dots, \mathcal{G}_N)$.  
Given an MRA task $\Phi^{[0,T]} = (0, T, \mathbb{T}, \mathbb{G})$, for any $0 \leq t_0 \leq T$, we define the truncated MRA task   as  
\[
\Phi_i^{[t_0,T]} = (t_0, T, \mathbb{T}_i, \mathbb{G}_i), 
\] 
which considers only the last $i$ target and safe regions with the start time shifted to $t_0$.

We first directly use result in \cite{fisac2015reach} to compute the feasible set of the task $\Phi_1^{[0,T]}=(0,T,\mathbb{T}_1,\mathbb{G}_1)$.
\begin{mylem}[\cite{fisac2015reach}]\label{lem:firstcase}
Let 
$h_{\mathsf{RA}}^{\Phi_1}(x,t)$  be the viscosity solution of HJ-VI in \eqref{eq:HJR} 
for target function $h_{\mathcal{T}}^{\Phi_1}$ and safe function $h_{\mathcal{G}}^{\Phi_1}$ defined by:  for any $(x,t) \in \mathbb{R}^n\times [0,T]$, we have\vspace{-6pt}
\begin{equation}
    \left\{
    \begin{array}{l} 
    h_{\mathcal{T}}^{\Phi_1}(x,t)=h_{\mathcal{T}_{N}}(x,t)\\
    h_{\mathcal{G}}^{\Phi_1}(x,t)= h_{\mathcal{G}_{N}}(x,t) \\
    \end{array}
\right..
\end{equation}
Then for any $(x,t) \in \mathbb{R}^n\times [0,T]$, it holds that
\begin{equation} \label{eq:firstcaseresult}
h_{\mathsf{RA}}^{\Phi_1}(x,t)\!\geq \!0 
\!  \Leftrightarrow \!
(\forall \gamma \in \Gamma_{[t,T]}, \exists \mathbf{u} \in \mathbb{U}_{[t,T]})[ \xi_{x,t}^{\mathbf{u},\gamma}\! \models \Phi^{[t,T]}_1].
\end{equation} 
\end{mylem}
Next, we prove that  the feasible set of MRA task $\Phi_{i+1}$ can be computed by regarding the feasible set of $\Phi_{i}$ as an additional time-varying target region.
\begin{mypro}\label{prop:secondcase}
For each $i \leq  N-1$, let $h_{\mathsf{RA}}^{\Phi_{i}}$ be a function such that, for any  $(x,t)\in \mathbb{R}^n\times [0,T]$, we have
\begin{equation}\label{eq:hypethesis}
    h_{\mathsf{RA}}^{\Phi_i}(x,t)\!\geq \! 0 \Leftrightarrow \! (\forall \gamma \!\in \!\Gamma_{[t,T]}, \exists \mathbf{u}\! \in \!\mathbb{U}_{[t,T]}) [\xi_{x,t}^{\mathbf{u},\gamma} \! \models \Phi^{[t,T]}_i]. 
\end{equation}
Let 
$h_{\mathsf{RA}}^{\Phi_{i+1}}(x,t)$ be the viscosity solution of HJ-VI in \eqref{eq:HJR} for target function $h_{\mathcal{T}}^{\Phi_{i+1}}$ and safe function $h_{\mathcal{G}}^{\Phi_{i+1}}(x,t)$  
defined by: for any $(x,t) \in \mathbb{R}^n\times [0,T]$, we have
\[
\left\{
    \begin{array}{l} 
    h_{\mathcal{T}}^{\Phi_{i+1}}(x,t)=\min\{ h_{\mathcal{T}_{N-i}}(x,t) ,h_{\mathsf{RA}}^{\Phi_{i}}(x,t)\}\\
    h_{\mathcal{G}}^{\Phi_{i+1}}(x,t)=h_{\mathcal{G}_{N-i}}(x,t) \\
    \end{array}
\right..
\]
Then for any $(x,t) \in \mathbb{R}^n\times [0,T]$, it holds that
\begin{equation} \label{eq:induction}
h_{\mathsf{RA}}^{\Phi_{i+1}}(x,t)\!\geq \!0 \! \Leftrightarrow \!
( \forall \gamma\! \in \!\Gamma_{[t,T]},\exists \mathbf{u} \!\in \!\mathbb{U}_{[t,T]})[ \xi_{x,t}^{\mathbf{u},\gamma}\! \models \!\Phi^{[t,T]}_{i+1}].
    \end{equation}
    \vspace{-2em}
\end{mypro}
\begin{pf}
Define sets $\mathcal{T}^\Phi_{i+1},\mathcal{G}^\Phi_{i+1} \subseteq \mathbb{R}^n \times [0,T]$ such that $(x,t) \in (\bullet)^\Phi_{i+1}\Leftrightarrow h_{(\bullet)}^{\Phi_{i+1}}(x,t) \geq 0$ with $(\bullet) \in \{\mathcal{T}, \mathcal{G}\}$. 
We prove that for any $(x,t) \in \mathbb{R}^n\times [0,T]$, it holds that
\begin{align} 
    &x \in\mathsf{RA}(t,T, \mathcal{T}^\Phi_{i+1}, \mathcal{G}^\Phi_{i+1}) \\
\Leftrightarrow \  & (\forall \gamma \in \Gamma_{[t,T]})(\exists \mathbf{u} \in \mathbb{U}_{[t,T]}) [\xi_{x,t}^{\mathbf{u},\gamma} \models \Phi^{[t,T]}_{i+1}].\nonumber 
\end{align}
$(\Rightarrow)$ 
Suppose that $x \in \mathsf{RA}(t,T, \mathcal{T}^\Phi_{i+1}, \mathcal{G}^\Phi_{i+1})$. Then for any $\gamma \in \Gamma_{[t,T]}$, we can find control function $\mathbf{u}\in \mathbb{U}_{[t,s]}$ with $s \in [t,T]$ s.t.
\begin{equation}\label{eq:secondmiddle4}
h_{\mathcal{T}}^{\Phi_{i+1}}\!(\xi_{x,t}^{\mathbf{u},\gamma}\!(s),s)\! \geq\! 0, \forall \tau\! \in \![t,s], h_{\mathcal{G}_{N-i}}(\xi_{x,t}^{\mathbf{u},\gamma}\!(\tau),\tau)\! \geq\! 0.
\end{equation}
From the definition of $h^{\Phi_{i+1}}_{\mathcal{T}}$, we have (a) $ h_{\mathcal{T}_{N-i}}(\xi_{x,t}^{\mathbf{u},\gamma}(s),s)$ $\geq 0$ and (b) $h^{\Phi_{i}}_\mathsf{RA}(\xi_{x,t}^{\mathbf{u},\gamma}(s),s) \geq 0$. From \eqref{eq:hypethesis} and (b) we can find control function $\mathbf{u}' \in \mathbb{U}_{[s,T]}$ such that
\begin{equation} \label{eq:secondmiddle1}
\xi_{x',s}^{\mathbf{u}',\gamma(\mathbf{u}')}\models \Phi^{[s,T]}_i, x' = \xi_{x,t}^{\mathbf{u},\gamma}(s).
\end{equation}
From \eqref{eq:secondmiddle1} and \eqref{eq:taskdes}, we can find 
$t_{N-i},t_{N-i+1},\dots,t_{N} \in [s,T]$ such that
\[
s = t_{N-i} \leq t_{N-i+1} \leq  \cdots \leq t_{N} \leq T,
\]
satisfying for any $N-i+1 \leq k \leq N$,
\begin{equation}\label{eq:secondmiddle6}
    \xi_{x',s}^{\mathbf{u}',\gamma(\mathbf{u}')}(t_k) \!\in\! \mathcal{T}_k(t_k),  \forall \tau \!\in\! [t_{k-1},t_k],  \xi_{x',s}^{\mathbf{u}',\gamma(\mathbf{u}')}(\tau)\! \in \!\mathcal{G}_k(\tau).
\end{equation}
Let $t_{N-i-1}=t$. By applying control function $\mathbf{u}'' \in \mathbb{U}_{[t,T]}$ s.t. $\mathbf{u}''(\tau) = \mathbf{u}(\tau)$ for $\tau \in [t,s)$ and $\mathbf{u}''(\tau) = \mathbf{u}'(\tau)$ for $\tau \in [s,T]$, from (a), \eqref{eq:secondmiddle4} and \eqref{eq:secondmiddle6}, for any $N-i\leq k \leq N$,
\begin{equation}\label{eq:secondmiddle7}
    \xi_{x,t}^{\mathbf{u}'',\gamma(\mathbf{u}'')}\!(t_k) \!\in \!\mathcal{T}_k(t_k), \forall \tau\! \in \![t_{k-1},t_{k}],  \xi_{x,t}^{\mathbf{u}'',\gamma(\mathbf{u}'')}\!(\tau)\! \in \!\mathcal{G}_k(\tau).
\end{equation}
From \eqref{eq:secondmiddle7} we know that $\xi^{\mathbf{u}'',\gamma(\mathbf{u}'')}_{x,t} \models \Phi_{i+1}^{[t,T]}$.
Proof from left hand side to right hand side is completed.

($\Leftarrow$) Suppose that for any $\gamma \in \Gamma_{[t,T]}$, we can find $\mathbf{u} \in \mathbb{U}_{[t,T]}$ such that $\xi_{x,t}^{\mathbf{u},\gamma} \models \Phi^{[t,T]}_{i+1}$. 
Then there are $t_{N-i-1},t_{N-i},\dots,t_{N} \in [t,T]$ such that
\[
t =  t_{N-i-1} \leq t_{N-i} \leq t_{N-i+1} \cdots \leq t_{N} \leq T
\]
satisfying 
for any $N-i \leq k \leq N$,
\begin{equation}\label{eq:secondmiddle8}
    \xi_{x,t}^{\mathbf{u},\gamma}(t_k) \in \mathcal{T}_k(t_k), \quad \forall \tau \in [t_{k-1},t_{k}],  \xi_{x,t}^{\mathbf{u},\gamma}(\tau) \in \mathcal{G}_k(\tau).
\end{equation}
It holds that
\[
\xi_{x',t_{N-i}}^{{\mathbf{u},\gamma}} \models  \Phi^{[t_{N-i},T]}_{i}, x'=\xi_{x,t}^{{\mathbf{u},\gamma}}(t_{N-i}).
\]
Then from \eqref{eq:hypethesis} we know that
\begin{equation} \label{eq:secondmiddle2}
    h^{\Phi_i}_{\mathsf{RA}}(\xi_{x,t}^{\mathbf{u},\gamma}(t_{N-i}),t_{N-i}) \geq 0.
\end{equation}
Since $t_{N-i}$ is arrival time of target $\mathcal{T}_{N-i}$, we know that
\begin{equation} \label{eq:secondmiddle3}
    h_{\mathcal{T}_{N-i}}(\xi_{x,t}^{\mathbf{u},\gamma}(t_{N-i}),t_{N-i}) \geq 0.
\end{equation}
From \eqref{eq:secondmiddle2} and \eqref{eq:secondmiddle3}, we have
\begin{equation}\label{eq:secondmiddle9}    h^{\Phi_{i+1}}_{\mathcal{T}}(\xi_{x,t}^{\mathbf{u},\gamma}(t_{N-i}),t_{N-i}) \geq 0.
\end{equation}
From \eqref{eq:secondmiddle8} we know that
\begin{equation}\label{eq:secondmiddle10}
\forall \tau \! \in\! [t,t_{N-i}], \!h^{\Phi_{i+1}}_{\mathcal{G}}\!(\xi_{x,t}^{\mathbf{u},\gamma}(\tau),\tau)\!=\! h_{\mathcal{G}_{N-i}}\!(\xi_{x,t}^{\mathbf{u},\gamma}(\tau),\tau)\! \geq \! 0.
\end{equation}
Thus $x \in \mathsf{RA}(t,T, \mathcal{T}^\Phi_{i+1}, \mathcal{G}^\Phi_{i+1})$ is true from \eqref{eq:defofHJRRA}, \eqref{eq:secondmiddle9} and \eqref{eq:secondmiddle10}. This completes the proof. $\hfill\square$
\end{pf}
Based on Proposition~\ref{prop:secondcase}, we immediately have the following result for the feasible set of the entire MRA task.
\begin{mythm} \label{thm:secondcombine}
    Let $h_{\mathsf{RA}}^{\Phi_{N}}(x,t)$ be the function in \eqref{eq:induction}.
    Then
\begin{equation}\label{eq:wanttoproveinT1}    h_{\mathsf{RA}}^{\Phi_{N}}(x,0)\geq 0 \Leftrightarrow x \in \mathsf{MRA}(0,T,\mathbb{T},\mathbb{G}).
    \end{equation}
\end{mythm}
\begin{remark} \label{remark:reasonforcontrol affine}\upshape 
In fact, the result of Theorem~\ref{thm:secondcombine} applies to general nonlinear systems without requiring the control-disturbance-affine assumption.
However, the control synthesis discussed in the next section involves optimization over control inputs and disturbances in real time, which may be impractical for systems without this assumption. 
Thus, when focusing solely on the feasible set of the MRA task, the system dynamics in \eqref{eq:systemdynamic} can be relaxed to $\dot{x}(t) = f(x,u,d,t)$, where $f:\mathbb{R}^n \times \mathcal{U} \times \mathcal{D} \times [0,T] \to \mathbb{R}^n$ is bounded uniformly continuous and Lipschitz continuous in $x$~\cite{fisac2015reach}.
\end{remark}

\section{Control Synthesis Procedure}\label{sec:controlsynthesis}
Suppose that  
the MRA task is feasible from the initial state $x_0 \in \mathbb{R}^n$, i.e.,  $x_0 \in \mathsf{MRA}(0,T,\mathbb{T},\mathbb{G})$. 
The objective of this section is to show how to explicitly compute state-feedback control function $\mathbf{c}:\mathbb{R}^n\times [0,T]\to \mathcal{U}$ to finish the MRA task $\Phi$ online.

Before going into technical details, we outline our control synthesis procedure, presented in Algorithm~\ref{alg:procedure}. The approach consists of the following key steps. 
First, in the offline computation stage, we compute the value functions 
\( h_{\mathsf{RA}}^{\Phi_i} \) for each subtask \( \Phi_i \), where \( i = 1, 2, \dots, N \).  
Then in the online execution stage,  suppose that the controller is during the execution of the \( i \)-th reach-avoid task, it ensures feasibility for subsequent tasks as follows:\vspace{-3pt}  
\begin{itemize}
    \item 
     \textbf{Current Functions Selection.}   
     The current value function is set to \( \mathtt{b} = h^{\Phi_{N-i+1}}_{\mathsf{RA}} \), which is the solution to HJ-VI defined in Proposition~\ref{prop:secondcase} (Line 3).   
     Then we 
     obtain the current target function \( h_{\mathcal{T}}^\mathtt{b} \) which is the dynamic target function of HJ-VI computing the current value function $\mathtt{b}$ (Line 4).  
     When this target function value is non-negative, the system is in target set and feasible set of future MRA task.\medskip
     \item 
     \textbf{Current Control Law Derivation.}   
     A time-varying state-feedback control law \( \mathbf{c}_\mathtt{b}(x,t) \) is derived from \( \mathtt{b} \), ensuring \( \mathtt{b}(x,t) \) remains non-negative over time. \medskip
     \item 
     \textbf{Termination and Transition.}   
      The control input \( \mathbf{c}_\mathtt{b}(x,t) \) is applied until \( h_{\mathcal{T}}^{\mathtt{b}}(x,t) \geq 0 \), indicating the current target is achieved and future task is feasible (Lines 6–7).  
      Once $h_{\mathcal{T}}^{\mathtt{b}}(x,t) \geq 0 $ is satisfied,  the target index \( i \) is updated in the for-loop, and the process repeats with the next value function until all targets are successfully reached.  
\end{itemize}

\begin{algorithm}[t]
\caption{Control Synthesis Procedure}\label{alg:procedure} 
\KwIn{Initial state $x_0 \in \mathbb{R}^n$ and value functions $h_{\mathsf{RA}}^{\Phi_i}$ for each $i=1,2,\dots,N$}

$x\gets x_0$, $t\gets 0$\;
\For{$i=1,2,\dots,N$}
{
set current value function by 
$\quad\quad \mathtt{b}(x,t)\gets h_{\mathsf{RA}}^{\Phi_{N-i+1}}(x,t)$\;
set current target function $h_{\mathcal{T}}^{\mathtt{b}}(x,t)$ by \eqref{eq:targetfunctionsynthesis}\; 
set state-feedback control law $\mathbf{c}_{\mathtt{b}}(x,t)$ by \eqref{eq:algorithmcontrolinputcondition}\;
\While{$ h_{\mathcal{T}}^{\mathtt{b}}(x,t) < 0$}
{
apply control input 
$\mathbf{c}_{\mathtt{b}}(x,t)$ and record new state $x$ and time $t$\;
}
}
\end{algorithm}

To be more specific, we first introduce the target function $h_{\mathcal{T}}^{\mathtt{b}}$ in line 4. As mentioned above, the current value function $\mathtt{b}$ is the solution of one of HJ-VIs computed in last section. Then the current target function $h_{\mathcal{T}}^{\mathtt{b}}$ is the dynamic target function of this HJ-VI. Specifically, for $\mathtt{b}=h_{\mathsf{RA}}^{\Phi_i}$ and $(x,t) \in \mathbb{R}^n \times [0,T]$, we define
\begin{equation}\label{eq:targetfunctionsynthesis}
   h_{\mathcal{T}}^{\mathtt{b}}(x,t)=
   \left\{
   \begin{array}{cl}
    h_{\mathcal{T}_{N-i+1}}(x)    & \text{if } i=1 \\
     \min\{ h_{\mathcal{T}_{N-i+1}}(x), h_{\mathsf{RA}}^{\Phi_{i-1}}(x,t) \}   & \text{otherwise}
   \end{array}.
   \right.
\end{equation} 
Now we explain how to derive the control function $\mathbf{c}_{\mathtt{b}}$ in line 5. We require the assumption as below.
\begin{assumption}
The value function $h_{\mathsf{RA}}^{\Phi_i}$  is differentiable over $\mathbb{R}^n \times [0,T]$ for $i=1,\dots,N$.  
\end{assumption}

\begin{remark}\upshape   
When the value functions associated with target $\mathcal{T}_i$ and constraint $\mathcal{G}_i$ are Lipschitz continuous, the reach-avoid value functions $h_{\mathsf{RA}}^{\Phi_i}(x,t)$ for $i = 1, 2, \dots, N$ inherit this Lipschitz continuity and are consequently differentiable almost everywhere. 
In practical implementations where the differential of $\mathtt{b}(x,t)$ may not exist at certain points, we follow the approach in \cite{9683085} by replacing the standard derivative in \eqref{eq:feasiblecontrolset} with either:
the \emph{superdifferential} (for non-smooth maximization problems), or  
the \emph{subdifferential} (for non-smooth minimization problems),  
as formally defined in \cite[Chapter 3.2.5]{bardi1997optimal}. This generalization enables practical control synthesis even at non-differentiable points. 
\end{remark}

For any value function $\mathtt{b}: \mathbb{R}^n \times [0,T] \to \mathbb{R}$, 
we define 
\begin{equation}\label{eq:feasiblecontrolset}
\begin{aligned}
&\mathtt{S}_{\mathtt{b}}(x,t)=\\
&\left\{\! u\! \in \!\mathcal{U} \!\,\middle\vert\, \! \!\frac{\partial \mathtt{b}(x,t)}{\partial x}\!(\!f(x,t) \!+\!g(x,t)u\!) 
  \!+\!p^\star\!(x,t)\!+\!\frac{\partial \mathtt{b}(x,t)}{\partial t} \!\geq\! 0\!\right\},
\end{aligned}
\end{equation}
where $p^\star(x,t)=\min_{d\in \mathcal{D}}\frac{\partial \mathtt{b}(x,t)}{\partial x}p(x,t)d$.
Clearly,  if we apply  control input from \eqref{eq:feasiblecontrolset}, then the time derivative of $\mathtt{b}$ satisfies
\begin{equation}\label{eq:neverdecrease}
    \dot{\mathtt{b}}(x(t),t) \!=\! \frac{\partial \mathtt{b}(x(t),t)}{\partial x}\!\Bar{f}(x(t),u(t),d,t) \!+ \frac{\partial \mathtt{b}(x(t),t)}{\partial t} \!\geq\! 0, 
\end{equation}
meaning $\mathtt{b}$ never decreases over time.
For technical purposes, we further define the \emph{feasible control input set}   by
\begin{equation} \label{eq:globalcontrol}
\Tilde{\mathtt{S}}_{\mathtt{b}}(x,t)\!= \!
		\left\{\!\!
		\begin{array}{cl}
			 \mathtt{S}_{\mathtt{b}}(x,t)  & \text{if }    h_{\mathcal{T}}^{\mathtt{b}}(x,t) < \mathtt{b}(x,t) \\
		 \mathcal{U}               & \text{otherwise}  
		\end{array}
		\right.\!\!\!.
\end{equation}
Now, let us  consider the case when the current target value function $h_{\mathcal{T}}^{\mathtt{b}}$ is negative and the current value function $\mathtt{b}$ is non-negative.  
We have the following two key observations:\vspace{-3pt}
\begin{itemize}
    \item 
    By adopting control input in $\mathtt{S}_{\mathtt{b}}$ defined in \eqref{eq:feasiblecontrolset},
    we can  ensure that the value of $\mathtt{b}$ never decreases over time. \medskip
    \item 
    Furthermore, the current target function will be non-negative at least at time $T$ since we have $h^{\mathtt{b}}_{\mathcal{T}}(x,T)\!=\!\mathtt{b}(x,T)\!\geq\! 0$ by boundary condition of HJ-VI. \vspace{-6pt}
\end{itemize}
By combining the above two observations together,  
we know that the desired control objective, i.e., current target is reached and future task is feasible, can be achieved by adopting feasible control input set in Eq.~\eqref{eq:globalcontrol}. This intuition above is formally stated as follow.
\begin{mypro}\label{prop:successreachtarget}
Given current value function $\mathtt{b}$ and current target function $h_{\mathcal{T}}^{\mathtt{b}}$, 
suppose that for $t_0 \in [0,T]$ and $x_0 \in \mathbb{R}^n$,
we have $\mathtt{b}(x_0,t_0) \geq 0 > h_{\mathcal{T}}^{\mathtt{b}}(x_0,t_0)$.
Let $\mathbf{c}_{\mathtt{b}}: \mathbb{R}^n\times[0,T]\to\mathcal{U}$ be a state-feedback control function which is uniformly continuous in $t$ and Lipschitz continuous in $x$, and
\begin{equation}\label{eq:algorithmcontrolinputcondition}
        \mathbf{c}_{\mathtt{b}}(x,t) \in \Tilde{\mathtt{S}}_{\mathtt{b}}(x,t), \forall x \in \mathbb{R}^n, t \in [0,T].
    \end{equation}
Then for any $\mathbf{d} \in \mathbb{D}_{[t_0,T]}$, we have
\begin{itemize}
    \item[(a)] $\exists t_1 \in [t_0,T]:h_{\mathcal{T}}^{\mathtt{b}}(\xi_{x_0,t_0}^{\mathbf{c}_{\mathtt{b}},\mathbf{d}}(t_1),t_1) \geq \mathtt{b}(\xi_{x_0,t_0}^{\mathbf{c}_{\mathtt{b}},\mathbf{d}}(t_1),t_1)$; \medskip
    \item[(b)]
    $\forall \tau \in [t_0,t_1]: \mathtt{b}(\xi_{x_0,t_0}^{\mathbf{c}_{\mathtt{b}},\mathbf{d}}(\tau),\tau) \geq 0 $.
\end{itemize}
\end{mypro}
\begin{pf}
We prove by contradiction that there exists $t_1 \in [t_0,T]$ such that
\begin{equation}\label{eq:positivemiddle}
    h_{\mathcal{T}}^{\mathtt{b}}(\xi_{x_0,t_0}^{\mathbf{c}_{\mathtt{b}},\mathbf{d}}(t_1),t_1) \geq \mathtt{b}(\xi_{x_0,t_0}^{\mathbf{c}_{\mathtt{b}},\mathbf{d}}(t_1),t_1).
\end{equation}
Assume that \eqref{eq:positivemiddle} is false.
Then $\mathbf{c}_{\mathtt{b}}(\xi^{\mathbf{c}_{\mathtt{b}},\mathbf{d}}_{x_0,t_0}(t),t) \in \mathtt{S}_{\mathtt{b}}(\xi^{\mathbf{c}_{\mathtt{b}},\mathbf{d}}_{x_0,t_0}(t),t)$ for $t\in [t_0,T]$.
From \eqref{eq:neverdecrease} we have $\dot{\mathtt{b}}(\xi^{\mathbf{c}_{\mathtt{b}},\mathbf{d}}_{x_0,t_0}(t),t) \geq 0$ for $t\in [t_0,T]$, i.e.,
\[
\mathtt{b}(\xi_{x_0,t_0}^{\mathbf{c}_{\mathtt{b}},\mathbf{d}}(\tau),\tau) \geq \mathtt{b}(\xi_{x_0,t_0}^{\mathbf{c}_{\mathtt{b}},\mathbf{d}}(t_0),t_0)\geq 0, \forall \tau \in [t_0,T].
\]
From the boundary condition of \eqref{eq:HJR}, it holds that
\[
h_{\mathcal{T}}^{\mathtt{b}}(\xi_{x_0,t_0}^{\mathbf{c}_{\mathtt{b}},\mathbf{d}}(T),T) \geq \mathtt{b}(\xi_{x_0,t_0}^{\mathbf{c}_{\mathtt{b}},\mathbf{d}}(T),T) \geq 0.
\]
It violates the assumption. Therefore, (a) is true. Moreover, we have $\dot{\mathtt{b}}(\xi^{\mathbf{c}_{\mathtt{b}},\mathbf{d}}_{x_0,t_0}(t),t) \geq 0$ for $t\in [t_0,t_1]$, i.e.,
\begin{equation}\label{eq:algcormiddle2}
\mathtt{b}(\xi_{x_0,t_0}^{\mathbf{c}_{\mathtt{b}},\mathbf{d}}(\tau),\tau) \geq 0, \forall \tau \in [t_0,t_1].
\end{equation}
Thus (b) holds. This completes the proof. $\hfill\square$
\end{pf}
One possible issue of feasible control input set $\Tilde{\mathtt{S}}_{\mathtt{b}}$ in \eqref{eq:globalcontrol} is that $\mathtt{S}_{\mathtt{b}}$ may be empty, which makes it impossible to construct the state-feedback control function $\mathbf{c}_{\mathtt{b}}$. However, the following result guarantees that such situation never happens.
\begin{mypro}\label{prop:feasiblecbf}
Given current value function $\mathtt{b}:\mathbb{R}^n \times [0,T]\to \mathbb{R}$,  
we have $\Tilde{\mathtt{S}}_{\mathtt{b}}(x,t) \neq \emptyset, \forall x \in \mathbb{R}^n, t \in [0,T]$.
\end{mypro}
\begin{pf}
It is sufficient to prove that
\begin{equation}\label{eq:feasiblemiddle2}
    h_{\mathcal{T}}^{\mathtt{b}}(x,t) < \mathtt{b}(x,t)  \implies \mathtt{S}_{\mathtt{b}}(x,t) \neq \emptyset .
\end{equation}
Consider $\mathtt{b}=h_{\mathsf{RA}}^{\Phi_{i}}$ for $i=1,2,\dots,N$. Since $h_{\mathsf{RA}}^{\Phi_{i}}(x,t)$ satisfies Equation \eqref{eq:HJR} and $\min\{ a,b\} = 0  \implies a \geq 0 $ , we know that
\begin{equation} \label{eq:feasiblemiddle1}
    \begin{aligned}
      \max \Big\{  \frac{\partial h_{\mathsf{RA}}^{\Phi_{i}}(x,t)}{\partial t} & \!+\!  \max_{u \in \mathcal{U} } \min_{d \in \mathcal{D}} \frac{\partial h_{\mathsf{RA}}^{\Phi_{i}}(x,t)}{\partial x} \Big(f(x,t)\!+\!g(x,t)u \\  
      &+p(x,t)d\Big), h_{\mathcal{T}}^{\Phi_{i}}(x,t)-h_{\mathsf{RA}}^{\Phi_{i}}(x,t) \Big\} \geq 0.
    \end{aligned}
\end{equation}
Combining with $h_{\mathcal{T}}^{\mathtt{b}}(x,t) < \mathtt{b}(x,t)$ and \eqref{eq:feasiblemiddle1}, we have
\[
   \frac{\partial h_{\mathsf{RA}}^{\Phi_{i}}(x,t)}{\partial t}+ \max_{u \in \mathcal{U} } \min_{d \in \mathcal{D}} \frac{\partial h_{\mathsf{RA}}^{\Phi_{i}}(x,t)}{\partial x} \Bar{f}(x,u,d,t)  \geq0.
\]
Thus \eqref{eq:feasiblemiddle2} holds. This completes the proof. $\hfill\square$
\end{pf}
\begin{remark}\upshape
In practice, given value function $\mathtt{b}$, time instant $t \in [0,T]$, and state $x \in \mathbb{R}^n$, when the control constraint set $\mathcal{U}$ is a polytope, the following quadratic programming (QP) problem can be solved to select the control input:  
\begin{equation} \label{eq:QP}  
    \begin{aligned}  
      &  \min_{u \in \mathcal{U}} \quad u^\top Q(x,t) u + F(x,t)^\top u \\  
      \text{s.t. } & \frac{\partial \mathtt{b}(x,t)}{\partial x}(f(x,t) \!+ \!g(x,t)u) + p^\star(x,t)+\frac{\partial \mathtt{b}(x,t)}{\partial t}\! \geq \!0,  
    \end{aligned}  
\end{equation}  
where $Q(x,t) \in \mathbb{R}^{m \times m}$ is a positive semi-definite matrix, $F(x,t) \in \mathbb{R}^m$, and $p^\star(x,t)=\min_{d\in \mathcal{D}}\frac{\partial \mathtt{b}(x,t)}{\partial x}p(x,t)d$. Under certain assumptions, as discussed in \cite{ames2016control,lindemann2018control}, the solution $u^\star(x,t)$ of QP \eqref{eq:QP} can be Lipschitz continuous in $x$ and uniformly continuous in $t$.  
Also, in many applications, a reference controller $u_{ref}(x,t): \mathbb{R}^n \times [0,T] \to \mathcal{U}$ is already provided, which may perform well on other criteria, such as energy efficiency. In such cases, the QP objective in \eqref{eq:QP} can be modified to $(u - u_{ref}(x,t))^\top Q (u - u_{ref}(x,t))$, ensuring minimal deviation from the reference controller while still satisfying the safety constraints. This approach allows the control input to meet the MRA task requirements while preserving the original performance as much as possible.
\end{remark}
Proposition~\ref{prop:successreachtarget} shows that each target can be visited under corresponding control law in line 5. 
Therefore, by an inductive argument, 
we show that Algorithm~\ref{alg:procedure} solves Problem~1  as formal statement below.
\begin{mythm}\label{thm:correctalgorithm}
Given system in \eqref{eq:systemdynamic},  MRA task $\Phi=(0,T,\mathbb{T},\mathbb{G})$, and initial state $x_0$ with  $x_0\in \mathsf{MRA}(0,T,\mathbb{T},\mathbb{G})$,
assume that $\mathbf{c}_{\mathtt{b}}$ in Proposition~\ref{prop:successreachtarget} can be found for each value function $\mathtt{b}$.
Then under any disturbance function $\mathbf{d} \in \mathbb{D}_{[0,T]}$, there exists $0 = \tau_0 \leq \tau_1 \leq \tau_2 \leq \cdots \leq \tau_N \leq T $ such that  Algorithm~\ref{alg:procedure} comes into $i$-th for-loop at time $\tau_{i-1}$ and MRA task $\Phi$ is finished at time $\tau_N$. 
\end{mythm}
\begin{pf}
From $x_0 \in \mathsf{MRA}(0,T,\mathbb{T},\mathbb{G})$ and Theorem~\ref{thm:secondcombine}, we have $h^{\Phi_N}_{\mathsf{RA}}(x_0,0)\geq 0$. 
Then under $\mathbf{c}_{\mathtt{b}}^1 $ in Proposition~\ref{prop:successreachtarget} for function $h^{\Phi_N}_{\mathsf{RA}}$ and $\mathbf{d} \in \mathbb{D}_{[0,T]}$, there is $\tau_1' \in [0,T]$ s.t.
\begin{align*}
   & h_{\mathcal{T}}^{\Phi_{N}}(\xi_{x_0,0}^{\mathbf{c}_{\mathtt{b}}^1,\mathbf{d}}(\tau_1'),\tau_1') \geq h^{\Phi_N}_{\mathsf{RA}}(\xi_{x_0,0}^{\mathbf{c}_{\mathtt{b}}^1,\mathbf{d}}(\tau_1'),\tau_1')\geq 0, \\
   & h_{\mathcal{G}}^{\Phi_{N}}(\xi_{x_0,0}^{\mathbf{c}_{\mathtt{b}}^1,\mathbf{d}}(\tau),\tau) \geq h^{\Phi_N}_{\mathsf{RA}}(\xi_{x_0,0}^{\mathbf{c}_{\mathtt{b}}^1,\mathbf{d}}(\tau),\tau) \geq 0,  \forall \tau \in [0,\tau_1'].
\end{align*}
Since trajectory $\xi_{x_0,0}^{\mathbf{c}_{\mathtt{b}}^1,\mathbf{d}}$ is continuous, the loop $i=1$ of Algorithm~\ref{alg:procedure} will terminal at $\tau_1 \in [0,\tau_1']$.
Now assume that at $\tau_k \in [0,T]$ and $x \in \mathbb{R}^n$ the Algorithm~\ref{alg:procedure} come into $(k+1)$-th loop.
Since $h^{\Phi_{N-k}}_{\mathsf{RA}}(x,\tau_k) \geq h^{\Phi_{N-k+1}}_{\mathcal{T}}(x,\tau_k) \geq 0$, 
under $\mathbf{c}_{\mathtt{b}}^{k+1}$ in Proposition~\ref{prop:successreachtarget} for function $h^{\Phi_{N-k}}_{\mathsf{RA}}$ and $\mathbf{d} \in \mathbb{D}_{[0,T]}$,
there is $\tau_{k+1}' \in [\tau_k,T]$ such that $\forall \tau \in [\tau_k,\tau_{k+1}']$,
\begin{align*}
   & h_{\mathcal{T}}^{\Phi_{N-k}}\!(\xi_{x,\tau_k}^{\mathbf{c}_{\mathtt{b}}^{k+1},\mathbf{d}}\!(\!\tau_{k+1}'\!),\!\tau_{k+1}'\!) \!\geq\!  h^{\Phi_{N-k}}_{\mathsf{RA}}\!(\xi_{x,\tau_k}^{\mathbf{c}_{\mathtt{b}}^{k+1},\mathbf{d}}\!(\!\tau_{k+1}'\!),\!\tau_{k+1}'\!)\!\geq\! 0, \\
   & h_{\mathcal{G}}^{\Phi_{N-k}}\!(\xi_{x,\tau_k}^{\mathbf{c}_{\mathtt{b}}^{k+1},\mathbf{d}}(\tau),\tau) \!\geq \!h^{\Phi_{N-k}}_{\mathsf{RA}}\!(\xi_{x,\tau_k}^{\mathbf{c}_{\mathtt{b}}^{k+1},\mathbf{d}}(\tau),\tau) \!\geq \!0.
\end{align*}
Then the $(k+1)$-th loop of Algorithm~\ref{alg:procedure} will terminal at $\tau_{k+1} \in [0,\tau_{k+1}']$.
Thus under $\mathbf{d} \in \mathbb{D}_{[0,T]}$, and state-feedback control function $\mathbf{c}:\mathbb{R}^n\times [0,T]\to \mathcal{U}$ such that $\mathbf{c}(x,\tau)=\mathbf{c}_{\mathtt{b}}^k(x,\tau)$ for $x \in \mathbb{R}^n$, $\tau \in [\tau_{k-1},\tau_k)$ and $k=1,\dots,N$, from the definition of functions $h_{\mathcal{T}}^{\Phi_{i}}$ and $h_{\mathcal{G}}^{\Phi_{i}}$ in Assumption~\ref{assum:global}, 
there exists $ 0 =  \tau_0  \leq  \tau_1   \leq \tau_2 \cdots \leq \tau_N \leq T $ such that, for any $i=1,\dots,N$, we have 
\begin{equation}
    \xi_{x_0,0}^{\mathbf{c},\mathbf{d}}(\tau_i) \in \mathcal{T}_i(\tau_i) \wedge \forall \tau \in [\tau_{i-1},\tau_i], \xi_{x_0,0}^{\mathbf{c},\mathbf{d}}(\tau) \in \mathcal{G}_{i}(\tau).
\end{equation}
This completes the proof. $\hfill\square$
\end{pf}
\begin{remark}\label{remark:modifycontrol}\upshape
The function $\mathtt{b}$ also represents the robustness of reaching a target while satisfying constraints. As previously discussed in \eqref{eq:neverdecrease}, the value of function $\mathtt{b}$ will never decrease over time. However, to allow for a larger admissible control set, we may relax this condition and only require $\mathtt{b}$ to remain above a threshold $\beta > 0$. In this case, the control set \eqref{eq:feasiblecontrolset} can be modified as  
\begin{equation}\label{eq:modifiedfeasiblecontrolset}  
\begin{aligned}  
    \mathtt{S}_{\mathtt{b}}^m(x,t) = \bigg\{ u &\in \mathcal{U} \;\bigg|\;  \frac{\partial \mathtt{b}(x,t)}{\partial x}(f(x,t) + g(x,t)u)+ \\  
    &p^\star(x,t) + \frac{\partial \mathtt{b}(x,t)}{\partial t} \geq -\alpha(\mathtt{b}(x,t) - \beta) \bigg\},  
\end{aligned}  
\end{equation}  
where $p^\star(x,t)=\min_{d\in \mathcal{D}}\frac{\partial \mathtt{b}(x,t)}{\partial x}p(x,t)d$ and $\alpha: \mathbb{R} \to \mathbb{R}$ is a strictly increasing, continuous function with $\alpha(0) = 0$. If $\mathtt{b}(x,t) - \beta \geq 0$, then $-\alpha(\mathtt{b}(x,t) - \beta) \leq 0$, ensuring $\mathtt{S}_{\mathtt{b}}(x,t) \subseteq \mathtt{S}_{\mathtt{b}}^m(x,t)$. Moreover, from \cite{ames2016control,lindemann2018control}, the control law derived by \eqref{eq:modifiedfeasiblecontrolset} guarantees $\mathtt{b}(x,t) \geq \beta$ for all time.  
\end{remark}
\section{Application to LTL Control Synthesis}\label{sec:discussall relation} 
Our model of multiple reach-avoid tasks is closely related to linear temporal logic (LTL), as both involve visiting regions with different properties in a specified order. However, synthesizing a controller for LTL tasks in general nonlinear systems with disturbances is an extremely challenging problem. In this section, we demonstrate that our method provides a sound, though not complete, approach to LTL control synthesis. 
\subsection{Co-Safe LTL and Finite-State Automata}
We consider the fragment of syntactically co-safe linear temporal logic without the next operator ($\text{scLTL}_{\setminus \bigcirc}$) with the following syntax\footnote{Similar to \cite{kloetzer2008fully,wongpiromsarn2015automata}, $\text{scLTL}_{\setminus \bigcirc}$ is chosen in this work because the system trajectory operates in continuous time, while the satisfaction of a formula is defined over discrete time.} 
\begin{equation}
    \varphi 
::= 
\text{True} 
\mid a 
\mid \neg a
\mid \varphi_1\vee \varphi_2
\mid \varphi_1\wedge \varphi_2
%\mid \bigcirc \varphi
\mid \varphi_1 U \varphi_2,
\end{equation}
where $a \in \mathcal{AP} $ is an atomic proposition;
$\neg$ and $\wedge$ are Boolean operators ``negation'' and ``conjunction'', respectively; $U$ is temporal operator ``until''. 

In general, LTL formulae are evaluated on infinite words over $2^{\mathcal{AP}}$. 
For infinite word $\rho\in (2^{\mathcal{AP}})^\omega$, we denote by $\rho\models \varphi$ if word $\rho$ satisfies LTL formula $\varphi$. The reader is referred to \cite{baier2008principles} for details of the semantics of LTL. 
However,  for a $\text{scLTL}_{\setminus \bigcirc}$ formula, its satisfaction can be determined in finite horizon. Specifically, for infinite word $\rho=\rho_{0}\rho_{1} \dots \in (2^{\mathcal{AP}})^\omega$ 
such that  $\rho\models \varphi$, it has a  \emph{finite good prefix} $\hat{\rho}=\rho_0\rho_1\dots \rho_n$ in the sense that $\hat{\rho}\rho'\models \varphi$ for any $\rho'\in  (2^{\mathcal{AP}})^\omega$.  
We denote by $\mathsf{Word}(\varphi)$ the set of all finite good prefixes for $\text{scLTL}_{\setminus \bigcirc}$ formula $\varphi$.  For a finite word $\rho\in (2^{\mathcal{AP}})^*$, we write $\rho \models \varphi$ if $\rho\in \mathsf{Word}(\varphi)$. 

The set of words satisfying a $\text{scLTL}_{\setminus \bigcirc}$ formula can be accepted by a (deterministic) \emph{finite state automata}  (FSA). Formally,  a FSA  is  a 5-tuple  
\begin{equation}
    A = (S,s_0,\Sigma,\delta,S_F),
\end{equation}
where $S$ is the set of states,  $s_0 \in S$ is the initial state,
$\Sigma$ is the alphabet,
$\delta: S \times \Sigma \rightarrow S$ is the deterministic partial transition function, 
and $S_F \subseteq S$ is the set of accepting states. 
The transition function  can be extended to $\delta: S\times \Sigma^{*}\to S$ recursively by: 
$\forall s \in S, \rho \in \Sigma^{*}, \sigma\in \Sigma, \delta(s,\rho\sigma)=\delta(\delta(s, \rho), \sigma)$ with $\delta(s,\epsilon)=s$. 
We denote by $\mathcal{L}(A)$ the set of all finite words \emph{accepted} by $A$, i.e., 
$\mathcal{L}(A)=\{\rho \in \Sigma^*:  \delta(s_0,\rho)\in S_F\}$.
For any scLTL formula  $\varphi$, there always exists a FSA $A_\varphi$ over $\Sigma = 2^{\mathcal{AP}}$ that only accepts all  good prefixes, i.e.,  $\mathcal{L}(A_\varphi) = \mathsf{Word}(\varphi)$~\cite{belta2017formal} 
.When LTL task $\varphi$ is without next operator, the accepting words of $\varphi$ is stutter-insensitive~\cite{baier2008principles}. That is, for FSA $A_\varphi$, for  any $s,s' \in S,\sigma \in \Sigma$, if $s=\delta(s',\sigma)$, we have $\delta(s,\sigma)=s$.

To specify the high-level property of the system trajectory,  let $L: \mathbb{R}^n \to 2^{\mathcal{AP}}$ be a labeling function for a finite set of atomic propositions $\mathcal{AP}$. 
We assume that  a new alphabet (set of atomic propositions) is generated whenever the system reaches a region with a different label than the previous one. Therefore, similar to  \cite{kloetzer2008fully, srinivasan2020control}, the word of a finite-time trajectory is defined as follows. 
\begin{mydef}[Trajectory Words]\label{def:trajectorylabel}\upshape
Let  $\xi: [t_0,t_1] \to \mathbb{R}^n$ be a trajectory and $L: \mathbb{R}^n \to 2^{\mathcal{AP}}$ be a labeling function. 
The word of trajectory $\xi$ under $L$, 
denoted by $L(\xi)$, 
is a sequence of sets of atomic propositions of form
\begin{equation}
    L(\xi)= l_0l_1\dots l_{n} \in (2^{\mathcal{AP}})^*
\end{equation}
such that
(i) $ l_{i-1} \neq l_{i},\forall i\leq n$; and 
(ii) there exists a sequence of time instants $t_0=a_0 < a_1< \dots<a_{n-1} < a_n \leq   t_1$ satisfying\vspace{-6pt}
    \begin{itemize}
        \item $L(\xi(a_{i}))=l_{i}$ for $i \leq n$;\vspace{3pt}
        \item $\forall \tau \in [a_n,t_1]$, $L(\xi(\tau))=l_n$.\vspace{3pt}
        \item for $i \leq n$, there exists $a_{i}' \in [a_{i-1},a_{i}]$ 
        with $L(\xi(a_{i}'))\in \{l_{i-1},l_i\}$ such that    $L(\xi(\tau))=l_{i-1},\forall \tau \in [a_{i-1}, a_{i}')$ 
            and  $L(\xi(\tau))=l_{i},\forall \tau \in (a_{i}', a_{i}]$. 
    \end{itemize}
\end{mydef} 
Note that we have excluded the trajectories generating infinite labels in finite horizon.
A trajectory $\xi$ satisfies $\text{scLTL}_{\setminus \bigcirc}$ task $\varphi$, denoted by $\xi\models \varphi$, if $L(\xi) \models \varphi$.  
 Our objective is still to find a control function such that the scLTL task is satisfied under any possible disturbances.  

\subsection{LTL Control Synthesis via MRA Tasks} 
The proposed MRA task can be used to enforce an LTL specification $\varphi$ based on its automata representation $A_\varphi$. The idea is to enforce the system trajectory to visit a sequence of states towards the accepting states. Such a sequence is referred to as a \emph{high-level plan}, defined as follows.

\begin{mydef}[High-Level Plans]\upshape
Let $\varphi$ be an LTL formula and $A_{\varphi}$ be its finite state automata (FSA). We call a sequence of states $\eta = s_0 s_1 \dots s_N$ a \emph{high-level plan} in $A_{\varphi}$, if: 
(i) $s_N \in S_F$, and
(ii) for all $i \leq N-1$, $\delta(s_i, \sigma) = s_{i+1}$ for some $\sigma \in \Sigma$; and
(iii) for all $i \leq N-1$, $s_i \neq s_{i+1}$. 
\end{mydef}

We say that a system trajectory $\xi$ follows a high-level plan $\eta = s_0 s_1 \dots s_N$ in $A_{\varphi}$, denoted by $\xi \models \eta$, if its word $L(\xi) = \sigma_1 \sigma_2 \dots \sigma_M$ traverses exactly through the states in $\eta$ in order (with possible loop stays at some states). That is, there exists a sequence of instants $0 = k_0 < k_1 < \dots < k_N \leq M$ with $k_{N+1}=M+1$ such that, for each $i \in {0,1,\dots,N}$, we have:
\begin{equation}\label{eq:highlevelplanswitch}
\forall j \in {k_{i}, \dots, k_{i+1}-1}: \delta(s_0, \sigma_1 \sigma_2 \cdots \sigma_j) = s_i.
\end{equation}
Clearly, according to this definition, if a trajectory $\xi$ follows a high-level plan in $A_{\varphi}$, then $\xi \models \varphi$.  We denote by $\mathsf{Fea}(t_0,t_1,\eta)$ the feasible set of task $\varphi$ following high level plan $\eta$ with start time $t_0$ and end time $t_1$, i.e.,
\begin{align}\label{eq:scltlhighlevelfeasible}  
&\mathsf{Fea}(t_0, t_1,\eta)\\
=
&\{x\in \mathbb{R}^n\mid\forall \gamma \in \Gamma_{[t_0,t_1]}, \exists \mathbf{u} \in \mathbb{U}_{[t_0,t_1]},   \xi_{x,t_0}^{\mathbf{u},\gamma}\models\eta \}.\nonumber
\end{align} 
Our approach is to convert a high-level plan following problem as a MRA task. 
To this end, recall that the execution of the system generates a single transition $\sigma \in 2^{\mathcal{AP}}$. 
Therefore, at each state $s_i$ in $\eta$, the following two cases are possible:\vspace{-6pt}
\begin{itemize}
\item 
The system makes a self-loop transition at state $s_i$, i.e., $\delta(s_i, \sigma) = s_i$;
\medskip
\item 
The system progresses towards the next state $s_{i+1}$, i.e., $\delta(s_i, \sigma) = s_{i+1}$.
\end{itemize}
These two cases correspond to the system moving to one of the following regions:\vspace{-3pt}
\begin{align}\label{eq:multiplereachavoidtransformone}
\mathcal{T}_i &= \{ x \in \mathbb{R}^n \mid \delta(s_{i-1}, L(x)) = s_i \}, \\
\mathcal{G}_i &= \{ x \in \mathbb{R}^n \mid \delta(s_{i-1}, L(x)) = s_{i-1} \}. \nonumber
\end{align}
Intuitively, at task stage $i$, the system needs to remain within the safe region $\mathcal{G}_i$, i.e., either contributing no new label or only the self-loop label at $s_{i-1}$, until it reaches the target region $\mathcal{T}_i$. Therefore, fulfilling the LTL task is equivalent to executing a high-level plan that leads to the accepting state, 
which is sufficient to fulfill the MRA task defined by $(\mathcal{T}_1, \dots, \mathcal{T}_N)$ and $(\mathcal{G}_1, \dots, \mathcal{G}_N)$.

However, the approach discussed above cannot be directly adopted due to the following issue. Since $\mathcal{T}_i$ and $\mathcal{G}_i$ correspond to different labeled regions, we have $\mathcal{T}_i \cap \mathcal{G}_i = \emptyset$. As a result, the system trajectory must cross the boundary of the safe region in order to reach the target region $\mathcal{T}_i$, which causes the current value function to decrease to $0$ during the control synthesis phase.

To address this challenge, our approach is to add state to safe region $\mathcal{G}_i$ and subtract state from target region $\mathcal{T}_i$ such that the new constructed sets $\Bar{\mathcal{G}}_i$ and $\Bar{\mathcal{T}}_i$ satisfy $\Bar{\mathcal{T}}_i\subseteq \Bar{\mathcal{G}}_i$.
Specifically, we proceed the following recursive construction.\vspace{-6pt}
\begin{itemize}
    \item 
    Initially, we consider the requirement of reaching the final accepting state in the high-level planning. 
    This can be easily implemented by taking the union of $\mathcal{G}_N$ and $\mathcal{T}_N$ as the enlarged safe region. Therefore, for each 
    $t \in [t_0,t_1]$, we define
    \begin{equation}\label{eq:intial recursive con}
        \Bar{\Phi}_1^{[t,t_1]}=(t,t_1,\Bar{\mathbb{T}}_{1}=(\mathcal{T}_N),\Bar{\mathbb{G}}_{1}=(\mathcal{G}_N\cup \mathcal{T}_N))
    \end{equation} 
    as the first time-varying MRA task constructed.
    \item 
    Now, suppose that, we have already obtained the   $(i-1)$th reach-avoid tasks 
    \begin{equation} 
         \Bar{\Phi}_{i-1}^{[t,t_1]}=(t,t_1,\Bar{\mathbb{T}}_{i-1},\Bar{\mathbb{G}}_{i-1}).  
    \end{equation}
    Then we define a new MRA task
    \begin{equation}\label{eq:recursively define task}
        \Bar{\Phi}_{i}^{[t,t_1]}=(t,t_1,
        \underbrace{(\Bar{\mathcal{T}}_{N-i+1},\Bar{\mathbb{T}}_{i-1})}_{=:\Bar{\mathbb{T}}_i},
        \underbrace{(\Bar{\mathcal{G}}_{N-i+1},\Bar{\mathbb{G}}_{i-1})}_{=:\Bar{\mathbb{G}}_i})
    \end{equation}
    where
    \begin{align}
\Bar{\mathcal{T}}_{N-i+1}(t)&= \mathsf{MRA}(\Bar{\Phi}_{i-1}^{[t,t_1]}) \cap \mathcal{T}_{N-i+1}(t), \label{eq:restriction}\\
\Bar{\mathcal{G}}_{N-i+1}(t)&= \mathcal{G}_{N-i+1}(t)\cup\Bar{\mathcal{T}}_{N-i+1}(t).
\end{align}
\end{itemize}
Intuitively, the new constructed target set $\Bar{\mathcal{T}}_{N-i+1}$ considers the feasibility of future MRA task $\Bar{\Phi}_{i-1}^{[t,t_1]}$.
Specifically, at initial construction in \eqref{eq:intial recursive con}, since there is no future task, the target set $\Bar{\mathcal{T}}_N$ is equal to the set $\mathcal{T}_N$.
Then, when constructing the MRA task $\Bar{\Phi}_{i}^{[t,t_1]}$ in \eqref{eq:recursively define task}, the target set $\Bar{\mathcal{T}}_{N-i+1}(t)$ is restricted on the $\mathsf{MRA}(\Bar{\Phi}_{i-1}^{[t,t_1]})$, i.e., the feasible set of future MRA task $\Bar{\Phi}_{i-1}^{[t,t_1]}$.
We discuss in Remark~\ref{remark:why} why such construction procedure is required when transforming high level plan following problem to MRA task.
 
The following result shows that the constructed MRA task $\Bar{\Phi}_{N}^{[t_0,t_1]}$ is appropriately defined. Specifically, the feasible set of LTL task $\varphi$ with high level plan $\eta$ in \eqref{eq:scltlhighlevelfeasible} is exactly the same as the feasible set of the MRA task $\Bar{\Phi}_{N}$ defined in \eqref{eq:recursively define task}.
For technical purpose, we assume without loss of generality,  that $\mathcal{T}_i$ in \eqref{eq:multiplereachavoidtransformone} is closed for any $i=1,2,\dots,N$. Otherwise, we can just consider its closure, which will not affect our result in practice.  

\begin{mypro}\label{prop:feasiblesetofMRAandLTLisequal}
Given dynamic system  \eqref{eq:systemdynamic} with initial state $x_0 \in \mathbb{R}^n$, 
$\text{scLTL}_{\setminus \bigcirc}$ task $\varphi$, a high level plan $\eta = s_0 s_1 \dots s_N$ in $A_\varphi$, let  
$\Bar{\Phi}_{N}^{[t_0,t_1]}$ be the MRA task  constructed according to $\eta$ as defined in  \eqref{eq:recursively define task}.  Then we have
\begin{equation}
    \mathsf{Fea}(t_0,t_1,\eta) = \mathsf{MRA}(\Bar{\Phi}_{N}^{[t_0,t_1]}).
\end{equation}
\end{mypro}
\begin{pf}
Define $[i]=\{1,2,\dots,i \}$ for a given integer $i$.
For state sets in \eqref{eq:multiplereachavoidtransformone}, 
since transition function of $A_\varphi$ is deterministic, $\mathcal{T}_i \cap \mathcal{G}_i=\emptyset$ for $i\in [N]$. 
We now prove by induction that for any $i\in [N]$, $t \in [t_0,t_1]$,
\begin{align}
        &\forall \gamma \in \Gamma_{[t,t_1]}, \exists \mathbf{u} \in \mathbb{U}_{[t,t_1]}, L(\xi_{x,t}^{\mathbf{u},\gamma})= l^{N-i+1}_1\cdots  l^{N-i+1}_{k_{N-i+1}} \nonumber \\
    & l^{N-i+2}_{1} \cdots   l^{N-1}_{k_{N-1}} l^N_1  \cdots l^N_{k_N} \wedge \Big( \forall m = N-i+1,\cdots   ,N, \nonumber  \\
    & \forall j\! \in \![k_{m}-1], \!\delta(s_{m-1}, l^{m}_{j})\!=\!s_{m-1} \wedge \!\delta(s_{m-1},l_{k_{m}}^{m})\!=\!s_m  \Big) \nonumber \\
    &\qquad \qquad \qquad\qquad\qquad\quad\Leftrightarrow x \in \mathsf{MRA} (\Bar{\Phi}_{i}^{[t,t_1]}).\label{eq:wanttoprove}
\end{align}
We provide some explanations on the notation. The superscript of label $l^m_n$ indicates that the high level plan is in $s_{m-1}$ and system is trying to reach $s_m$. The subscript is the index for label sequence to reach $s_m$ from $s_{m-1}$ with length $k_{m}$. 
When $i=1$, for $t\in [t_0,t_1]$, $\gamma \in \Gamma_{[t,t_1]}$, and $\mathbf{u} \in \mathbb{U}_{[t,t_1]}$, we have
\begin{align*}
     &L(\xi_{x,t}^{\mathbf{u},\gamma})= l^N_1\dots l^N_{k_N}\wedge \delta(s_{N-1},l_{k_{N}}^{N})=s_N\wedge\\
    & \quad \forall j \in [k_{N}-1], \delta(s_{N-1},l^{N}_{j})=s_{N-1}\\
    \Leftrightarrow & \exists t^N \in [t,t_1], \xi_{x,t}^{\mathbf{u},\gamma}(t^N) \in \mathcal{T}_N(t^N) \wedge \\
    & \quad \qquad \forall \tau \in [t,t^N),\xi_{x,t}^{\mathbf{u},\gamma}(\tau) \in \mathcal{G}_N(\tau)  \\
      \Leftrightarrow & \exists t =  \tau_{N-1}  \leq  \tau_N \leq t_1 \text{ s.t. }\xi_{x,t}^{\mathbf{u},\gamma}(\tau_N) \in \mathcal{T}_N(\tau_N) \wedge\\ 
      & \qquad  \forall \tau \in [\tau_{N-1},\tau_N], \xi_{x,t}^{\mathbf{u},\gamma}(\tau) \in \mathcal{G}_N(\tau) \cup \mathcal{T}_N(\tau). \nonumber
\end{align*}
The first ``$\Leftrightarrow$'' holds since (a) $\mathcal{T}_N$ is closed and $\mathcal{T}_N \cap \mathcal{G}_N=\emptyset$, and thus the trajectory will be in $\mathcal{T}_N$ when crossing the boundary of $\mathcal{T}_N$ and $\mathcal{G}_N$, and (b) the sets $\mathcal{G}_N$ and $\mathcal{T}_N$ in \eqref{eq:multiplereachavoidtransformone} exactly include state with corresponding labels.
The ``$\Leftarrow$'' of second ``$\Leftrightarrow$'' holds since $\mathcal{T}_N$ is closed and there is $\tau' \in [\tau_{N-1},\tau_N]$ s.t. $\forall \tau \in [\tau_{N-1},\tau'), \xi_{x,t}^{\mathbf{u},\gamma}(\tau) \in \mathcal{G}_N(\tau) \wedge \xi_{x,t}^{\mathbf{u},\gamma}(\tau') \in \mathcal{T}_N(\tau')$. 
From \eqref{eq:taskdes} and \eqref{eq:MRAfeasible}, we know \eqref{eq:wanttoprove} holds for $i=1$. Assume that \eqref{eq:wanttoprove} is true for $n=i$. Now consider the case $n=i+1>1$. For $t\in [t_0,t_1]$, $\gamma \in \Gamma_{[t,t_1]}$, and $\mathbf{u} \in \mathbb{U}_{[t,t_1]}$, we have
\begin{align*}
    &  L(\xi_{x,t}^{\mathbf{u},\gamma})= l^{N-i}_1\cdot \cdot l^{N-i}_{k_{N-i}}l^{N-i+1}_{1} \cdot \cdot   l^{N-1}_{k_{N-1}} l^N_1\cdot \cdot l^N_{k_N} \\
    & \wedge \bigg( \forall m = N-i,\dots,N, \delta(s_{m-1},l_{k_{m}}^{m})=s_m \wedge \\
    & \qquad \qquad \forall j \in [k_{m}-1], \delta(s_{m-1},l^{m}_{j})=s_{m-1} \bigg)\\
     \Leftrightarrow   &  L(\xi_{x,t}^{\mathbf{u},\gamma})= l^{N-i}_1\cdot \cdot l^{N-i}_{k_{N-i}}l^{N-i+1}_{1} \cdot \cdot   l^{N-1}_{k_{N-1}} l^N_1\cdot \cdot l^N_{k_N}   \\
     & \wedge \forall j \in [k_{N-i}-1], \delta(s_{N-i-1},l^{N-i}_{j})=s_{N-i-1}  \\
    & \wedge \delta(s_{N-i-1},l_{k_{N-i}}^{N-i})=s_{N-i} \wedge \bigg (\exists t^{N-i} \in [t,t_1], x'=\\
&\xi_{x,t}^{\mathbf{u},\gamma}(t^{N-i}) \wedge L(x')=l^{N-i}_{k_{N-i}} \wedge x' \in \mathsf{MRA}(\Bar{\Phi}_{i}^{[t^{N-i},t_1]}) \bigg )\\
    \Leftrightarrow & \exists t^{N-i} \in [t,t_1], \xi_{x,t}^{\mathbf{u},\gamma}(t^{N-i}) \in \mathcal{T}_{N-i}(t^{N-i}) \\
    & \quad  \wedge\forall \tau \in [t,t^{N-i}),\xi_{x,t}^{\mathbf{u},\gamma}(\tau) \in \mathcal{G}_{N-i}(\tau)  \\
    & \quad  \wedge \xi_{x,t}^{\mathbf{u},\gamma}(t^{N-i}) \in \mathsf{MRA}(\Bar{\Phi}_{i}^{[t^{N-i},t_1]}) \\
      \Leftrightarrow & \exists t =  \tau_{N-i-1}  \leq  \tau_{N-i} \leq t_1 \text{ s.t. } \\
      &\xi_{x,t}^{\mathbf{u},\gamma}(\tau_{N-i}) \in \mathcal{T}_{N-i}(\tau_{N-i}) \cap \mathsf{MRA}(\Bar{\Phi}_{i}^{[\tau_{N-i},t_1]}) \\ 
      &\wedge \forall \tau \in [\tau_{N-i-1},\tau_{N-i}], \xi_{x,t}^{\mathbf{u},\gamma}(\tau) \in \Bar{\mathcal{G}}_{N-i}(\tau).
\end{align*}
The first ``$\Leftrightarrow$'' comes from (a) $\delta(s_{N-i},l^{N-i}_{k_{N-i}})=s_{N-i}$ since $\delta(s_{N-i-1},l^{N-i}_{k_{N-i}})=s_{N-i}$, and (b) \eqref{eq:wanttoprove} is true for $n=i$.
The ``$\Rightarrow$'' of second ``$\Leftrightarrow$'' holds since (a) $\mathcal{T}_{N-i}$ is closed and $\mathcal{T}_{N-i} \cap \mathcal{G}_{N-i}=\emptyset$, and thus the trajectory will be in $\mathcal{T}_{N-i}$ when crossing the boundary of $\mathcal{T}_{N-i}$ and $\mathcal{G}_{N-i}$, and (b) the sets $\mathcal{G}_{N-i}$ and $\mathcal{T}_{N-i}$ in \eqref{eq:multiplereachavoidtransformone} exactly include state with corresponding labels.
We now explain why the ``$\Leftarrow$'' of last ``$\Leftrightarrow$'' is true.
Since $\mathcal{T}_{N-i}$ is closed, there is $\tau' \in [\tau_{N-i-1},\tau_{N-i}]$ s.t. $\forall \tau \in [\tau_{N-i-1},\tau'), \xi_{x,t}^{\mathbf{u},\gamma}(\tau) \in \mathcal{G}_{N-i}(\tau) \wedge \xi_{x,t}^{\mathbf{u},\gamma}(\tau') \in \mathcal{T}_{N-i}(\tau')$.
That is, $\tau'$ is the first time trajectory reaches target $\mathcal{T}_{N-i}$ during $[\tau_{N-i-1},\tau_{N-i}]$. Since $\xi_{x,t}^{\mathbf{u},\gamma}(\tau') \in \Bar{\mathcal{G}}_{N-i}(\tau')$ and $\mathcal{G}_{N-i}\cap \mathcal{T}_{N-i}=\emptyset$, we have $\xi_{x,t}^{\mathbf{u},\gamma}(\tau') \in \mathcal{T}_{N-i}(\tau') \cap \mathsf{MRA}(\Bar{\Phi}_i^{[\tau',t_1]})$. Thus ``$\Leftarrow$'' of third ``$\Leftrightarrow$'' holds.
Then from \eqref{eq:taskdes} and \eqref{eq:MRAfeasible}, \eqref{eq:wanttoprove} holds for $n=i+1$.
Combining \eqref{eq:wanttoprove} with $i=N$, $t=t_0$ and \eqref{eq:highlevelplanswitch}, we have
\[
\begin{aligned}
    \forall \gamma \in \Gamma_{[t_0,t_1]},\exists \mathbf{u} \in \mathbb{U}_{[t_0,t_1]}, &L(\xi_{x,t_0}^{\mathbf{u},\gamma})\models\eta  \\
    &\Leftrightarrow x \in \mathsf{MRA}(\Bar{\Phi}_{N}^{[t_0,t_1]}).
\end{aligned}
\]
Finally combining with \eqref{eq:scltlhighlevelfeasible} we complete the proof.
$\hfill\square$
\end{pf}
The following theorem further establishes the soundness of our approach, which states that 
in order to ensure the satisfaction of the LTL task $\varphi$, it suffices to run Algorithm~\ref{alg:procedure} to ensure the MRA task $\Bar{\Phi}_{N}^{[t_0,t_1]}$ constructed from a high-level plan $\eta$ in $A_\varphi$. 
\begin{mythm}\label{thm:highlevelmultiplereachequivalent}
Given  system dynamic \eqref{eq:systemdynamic} with initial state $x_0 \in \mathbb{R}^n$,  
$\text{scLTL}_{\setminus \bigcirc}$ task $\varphi$, 
a high level plan $\eta = s_0 s_1 \dots s_N$ in $A_\varphi$, 
let  
$\Bar{\Phi}_{N}^{[0,T]}$ be the MRA task  constructed according to $\eta$ as defined in  \eqref{eq:recursively define task}. 
Suppose that all conditions in Theorem~\ref{thm:secondcombine} and \ref{thm:correctalgorithm} hold for the MRA task $\Bar{\Phi}_{N}^{[0,T]}$.
Then for any possible outcome trajectory $\xi$ of Algorithm~\ref{alg:procedure}, we have $\xi\models \varphi$.
\end{mythm}
\begin{pf}
For any possible outcome trajectory $\xi$ of Algorithm~\ref{alg:procedure},
let
\[
0 = \tau_0 \leq \tau_1 \leq \tau_2 \leq \cdots \leq \tau_N \leq T  ,
\]  
be a sequence of time instants such that for-loop of Algorithm~\ref{alg:procedure} comes from target $\Bar{\mathcal{T}}_{i}$ to target $\Bar{\mathcal{T}}_{i+1}$ at time $\tau_{i}$ for $i=1,2,\dots,N-1$.
From Theorem~\ref{thm:correctalgorithm}, it also holds that for $i=1,2,\dots,N$,
\begin{equation}
 \bigg[\xi(\tau_i) \in \Bar{\mathcal{T}}_i(\tau_i)\bigg]  
\wedge  
\bigg[\forall \tau \in [\tau_{i-1}, \tau_i]: \xi(\tau) \in \Bar{\mathcal{G}}_i(\tau) \bigg].  \nonumber
\end{equation}
We now prove by contradiction that
\begin{equation}\label{eq:lastmiddle}
    \xi(\tau) \in \mathcal{G}_i(\tau), \quad \forall i = 1,2,\dots,N, \tau \in [\tau_{i-1},\tau_i).
\end{equation}
Assume that there exists $\tau_i' \in [\tau_{i-1},\tau_i)$ such that $\xi(\tau_i') \in \Bar{\mathcal{T}}_i(\tau_i')$ for some $i=1,2,\dots,N$.
since $\Bar{\mathcal{T}}_i$ in \eqref{eq:restriction} has already consider the feasibility of future MRA task $\Bar{\Phi}_{N-i}^{[t_0,t_1]}$, $h_{\mathcal{T}}^{\mathtt{b}_i}$ in \eqref{eq:targetfunctionsynthesis} satisfies that $h_{\mathcal{T}}^{\mathtt{b}_i}(\xi(\tau_i'),\tau_i')\geq 0$.
Thus the for-loop of the Algorithm~\ref{alg:procedure} will switch to target $\Bar{\mathcal{T}}_{i+1}$ at $\tau_{i}'$,
violating the condition that Algorithm~\ref{alg:procedure} switches to $\Bar{\mathcal{T}}_{i+1}$ until $\tau_{i}$.
From \eqref{eq:lastmiddle} and \eqref{eq:multiplereachavoidtransformone}, we have
\begin{align}
    &\delta(s_{i-1},L(\xi(\tau)))=s_{i-1}, \forall \tau \in [\tau_{i-1},\tau_i), \label{eq:inrightlabel}\\
    &\delta(s_{i-1},L(\xi(\tau_i)))=s_i, \quad \forall i=1,2,\dots, N. \label{eq:reachtarget}
\end{align}
Finally, from \eqref{eq:inrightlabel}, \eqref{eq:reachtarget} and \eqref{eq:highlevelplanswitch}, we get $\xi\models\varphi$. $\hfill\square$
\end{pf}
\begin{remark}\upshape
Our approach is sound in the sense that if the heuristically selected high-level plan can be followed, then the LTL task is successfully enforced. However, if the selected plan cannot be realized, one needs to select an alternative plan. Yet, in general, there exist infinitely many high-level plans. A practical solution is to enumerate all high-level plans within a bounded length. In particular, when the FSA $A_\varphi$ contains no cycles other than self-loops, the maximum length of high-level plan never exceeds the size of the state set of $A_\varphi$, and in practice problems of interest should also be solvable within such a finite bound. In this case, such enumeration provides a solution to the LTL task control synthesis that is both sound and complete.
\end{remark}
 
\begin{remark}\label{remark:why}\upshape 
Finally, we remark that the satisfaction of the MRA task $\Bar{\Phi}_{N}^{[t_0,t_1]}$ does not directly imply the satisfaction of the LTL task $\varphi$ under the high-level plan $\eta$. For instance, consider a trajectory that remains within $\Bar{\mathcal{G}}_i$ and reaches $\Bar{\mathcal{T}}_i$ multiple times during this interval.
According to the definition in \eqref{eq:taskdes}, such a trajectory still satisfies the MRA task $\Bar{\Phi}_{N}^{[t_0,t_1]}$, since $\Bar{\mathcal{T}}_i \subseteq \Bar{\mathcal{G}}_i$ from \eqref{eq:restriction}, and we may select $\tau_i$ in \eqref{eq:taskdes} as the last arrival time of $\Bar{\mathcal{T}}_i$.
However, this trajectory may violate the high-level plan. Specifically, once the target $\Bar{\mathcal{T}}_i$ is first reached, a new label is generated, and the high-level plan transitions to the next automata state according to \eqref{eq:highlevelplanswitch}.
Consequently, the segment between the first and last visits to $\Bar{\mathcal{T}}_i$ may yield undesired labels for the high-level plan.
Interestingly, Algorithm~\ref{alg:procedure} prevents this issue by switching to the for-loop of the next target immediately after the current target is completed. This is also why the MRA task $\Bar{\Phi}_{N}^{[t_0,t_1]}$ must be constructed via the recursive procedure in \eqref{eq:recursively define task}.
In particular, the construction ensures that the current target accounts for the feasibility of future tasks.
Without this consideration, it would be unreasonable to switch to the next target solely upon completing the current one.
\end{remark} 
\section{Case Studies and Simulations} \label{sec:simulation}
\begin{figure}[tp]
    \subfigure[Workspace.] 
	{\label{fig:workspace1}
 \begin{minipage}[b]{0.48\linewidth}
	\centering
 \includegraphics[height=4.2cm]{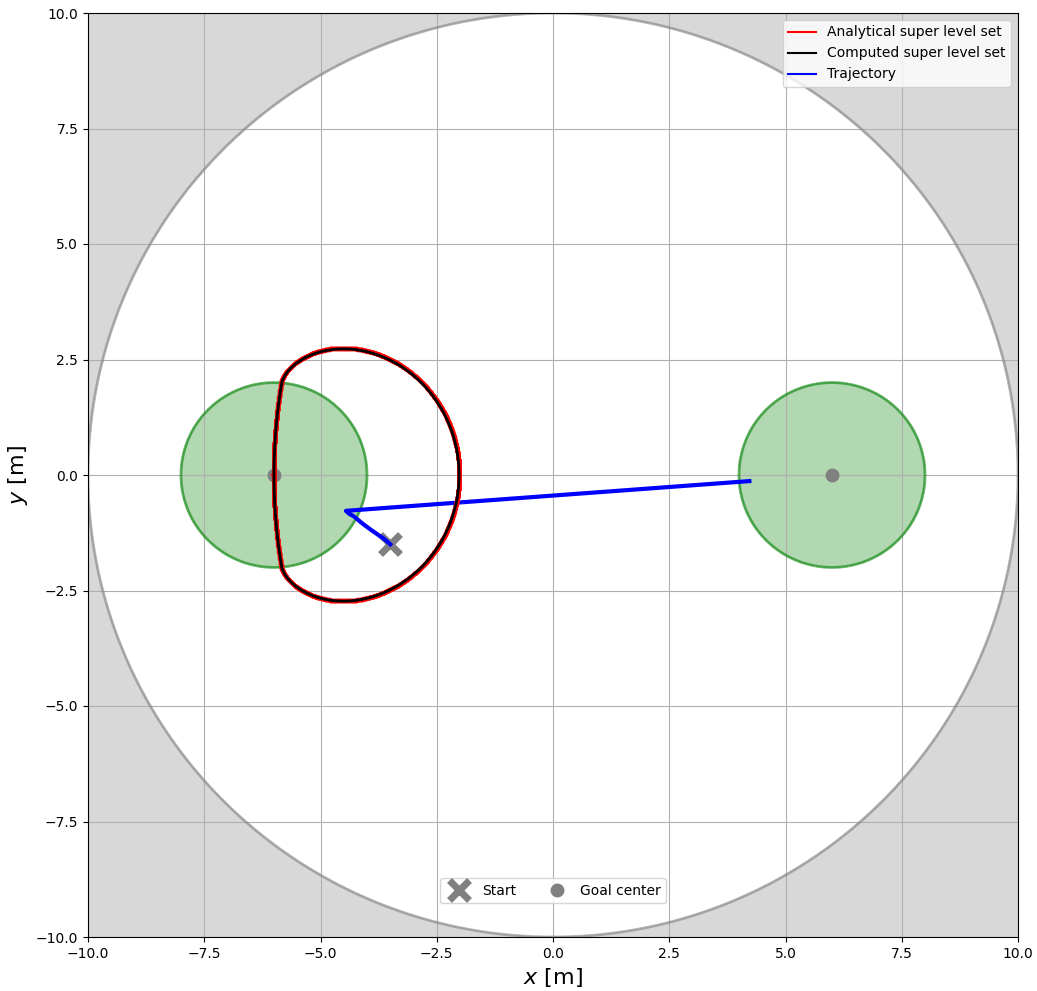}
    \end{minipage}
	}\subfigure[Part of FSA of task $\varphi$.] 
	{\label{fig:DFAcase1}
 \begin{minipage}[b]{0.48\linewidth}
	\centering
 \begin{tikzpicture}
[
acc/.style={circle, draw=black!255, fill=white!255, very thick, double,minimum height=6mm,minimum width=8mm,rounded corners = 1mm},
nor/.style={circle, draw=black!255, fill=white!255, very thick, minimum height=6mm,minimum width=6mm},
]
	\node [nor](q1)at(0,0){$s_{0}$};
	\node [nor](q2)at(2.5,0){$s_{2}$};
	\node [nor](q3)at(0,2){$s_{1}$};
	\node [acc](q4)at(2.5,2){$s_{F}$};

\draw[-{Stealth}] (-0.8,0) -- (-0.35,0);
%\draw[-{Stealth}] (0,0.35) -- (0,0.7)-- (4,0.7)--(4,0.3);
\draw[-{Stealth}] (q1) -- (q3);
\draw[-{Stealth}] (q1) -- (q2);
\draw[-{Stealth}] (q2) -- (q4);
\draw[-{Stealth}] (q3) -- (q4);
\draw[-{Stealth}] (q1) .. controls +(right:10mm) and +(down:10mm) .. (q1);
\draw[-{Stealth}] (q2) .. controls +(right:10mm) and +(down:10mm) .. (q2);
\draw[-{Stealth}] (q3) .. controls +(left:10mm) and +(up:10mm) .. (q3);
\draw[-{Stealth}] (q4) .. controls +(right:10mm) and +(up:10mm) .. (q4);

\node (c33)at(0.45,2.6){$\neg a \wedge e$};
\node (c12)at(1.25,0.2){$a \wedge \neg b \wedge e$};
\node (c23)at(2.1,0.9){$ b$};
\node (c22)at(2.2,-0.7){$\neg b \wedge e$};
\node (c11)at(0.55,-0.7){$\neg a \wedge \neg b \wedge e$};
\node (c34)at(1.25,2.2){$a$};
\node (c44)at(3.1,2.55){$1$};
\node (c13)at(-0.25,0.9){$\neg a \wedge b \wedge e$};
\end{tikzpicture}
    \end{minipage}
	}
    \caption{Simulation Result for the Case of Single Integrators.}
   \label{fig:case1}
\end{figure}
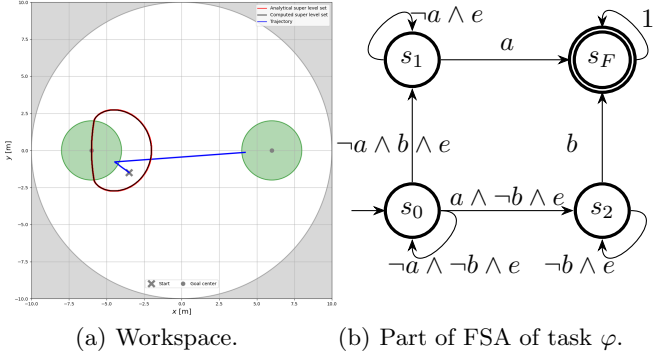
We have implemented our proposed value function computation and control synthesis procedure in Python. The HJR PDE is solved using $\textsf{JAX}$ by dynamic programming method which suffers from curse of dimensionality and can only be applied to system with dimension smaller than $6$. However, dynamic programming method can use GPU for acceleration, which to some extent alleviates this problem. In this section, we illustrate our algorithm by applying to four case studies: single integrators, double integrators, spacecraft
rendezvous and kinematic unicycles robots. 
For all examples, the offline value function computation takes only a few seconds on a RTX-3090 desktop. 

\subsection{Single Integrators}
We  consider a  mobile robot modeled by a single-integrator  
\begin{equation} \label{eq:single-integrator dynamics}
    \dot{x} = u,
\end{equation}
where $x=(x_1,x_2) \in \mathbb{R}^2$ and $u=(u_1,u_2) \in \mathcal{U} \subseteq \mathbb{R}^2$ such that $\mathcal{U}=\{ u \in \mathbb{R}^2 \mid \Vert u \Vert_2 \leq 1 \}$. We define regions of interest by $R_1=\{ x \in \mathbb{R}^2 \mid (x_1+6)^2 + x_2^2 \leq 2^2 \}$, $R_2=\{ x \in \mathbb{R}^2 \mid (x_1-6)^2 + x_2^2 \leq 2^2 \}$,
$R_3=\{ x \in \mathbb{R}^2 \mid x_1^2 + (x_2+6)^2 \leq 2^2 \}$, $R_4=\{ x \in \mathbb{R}^2 \mid x_1^2 + (x_2-6)^2 \leq 2^2 \}$, and $R_5=\{ x \in \mathbb{R}^2 \mid x_1^2 + x_2^2 \leq 10^2 \}$ and assign label $a,b,c,d,e$, respectively, to states in these regions.
We will omit the time variable for target and safe regions when they are static.
The scLTL task   is
\[
\varphi =  ( e \text{ }U a \wedge e \text{ }U b )\vee ( e \text{ }U c \wedge e \text{ }U d ),
\]
i.e., the robot should reach either $R_1$ and $R_2$ or $R_3$ and $R_4$ while always staying in $R_5$. We convert formula $\varphi$ to FSA whose partial structure is shown in Figure~\ref{fig:DFAcase1}. 
All transitions in FSA, except the transitions towards state $s_F$, should add condition $\neg c \wedge \neg d$, which is omitted for simplicity. 
We choose the high-level plan as $\eta=s_0s_2s_F$. 
Therefore, the target and safe regions in \eqref{eq:multiplereachavoidtransformone} are defined by $\mathcal{T}_1=R_1$, $\mathcal{T}_2=R_2$, $\mathcal{G}_1=R_5\setminus (R_1 \cup R_2\cup R_3\cup R_4)$ and $\mathcal{G}_2=R_5 \setminus (R_2 \cup R_3 \cup R_4)$. The start time is $0$ and the end time is $10$.
We use signed distance function as value function of each target region and safe region. For example, for target $R_1$, we define 
\begin{align}
h_{R_1}(x,t) \!= \!
		\left\{\!\!
		\begin{array}{cl}
			\sqrt{2^2 - (x_1+6)^2 - x_2^2}  &    x \in R_1, t \in [0,10] \\
		-\sqrt{(x_1+6)^2 + x_2^2-2^2}               & x \notin R_1, t \in [0,10]  
		\end{array}
		\right.\!\!\!. \nonumber
\end{align}
The workspace for this case study is shown in Figure~\ref{fig:workspace1}. The numerically computed feasible set of the MRA task, derived from Theorem~\ref{thm:secondcombine}, is represented by the black line. Due to the simplicity of the system dynamics and the MRA task, we also analytically compute the feasible set, depicted by the red line in Figure~\ref{fig:workspace1}. The two sets overlap precisely, which further validates the correctness of Theorem~\ref{thm:secondcombine}.

For all case studies, we use a system sampling time of 0.01s
 and the control input is computed by solving Program \eqref{eq:QP} with zero-order hold and a control update period of 0.1s.
 In this case study, the average solving time for \eqref{eq:QP} is 0.009s, making it sufficiently fast for online implementation. The robot's initial state is set to \( x_0 = (-3.5, -1.5) \), and Figure~\ref{fig:workspace1} displays the simulation trajectory, demonstrating successful completion of the MRA task and LTL task. 
\begin{figure*}[tp]
	\subfigure[Double Integrators.] 
	{\label{fig:case2}
 \begin{minipage}[b]{0.33\linewidth}
	\centering
 \includegraphics[height=5.8cm]{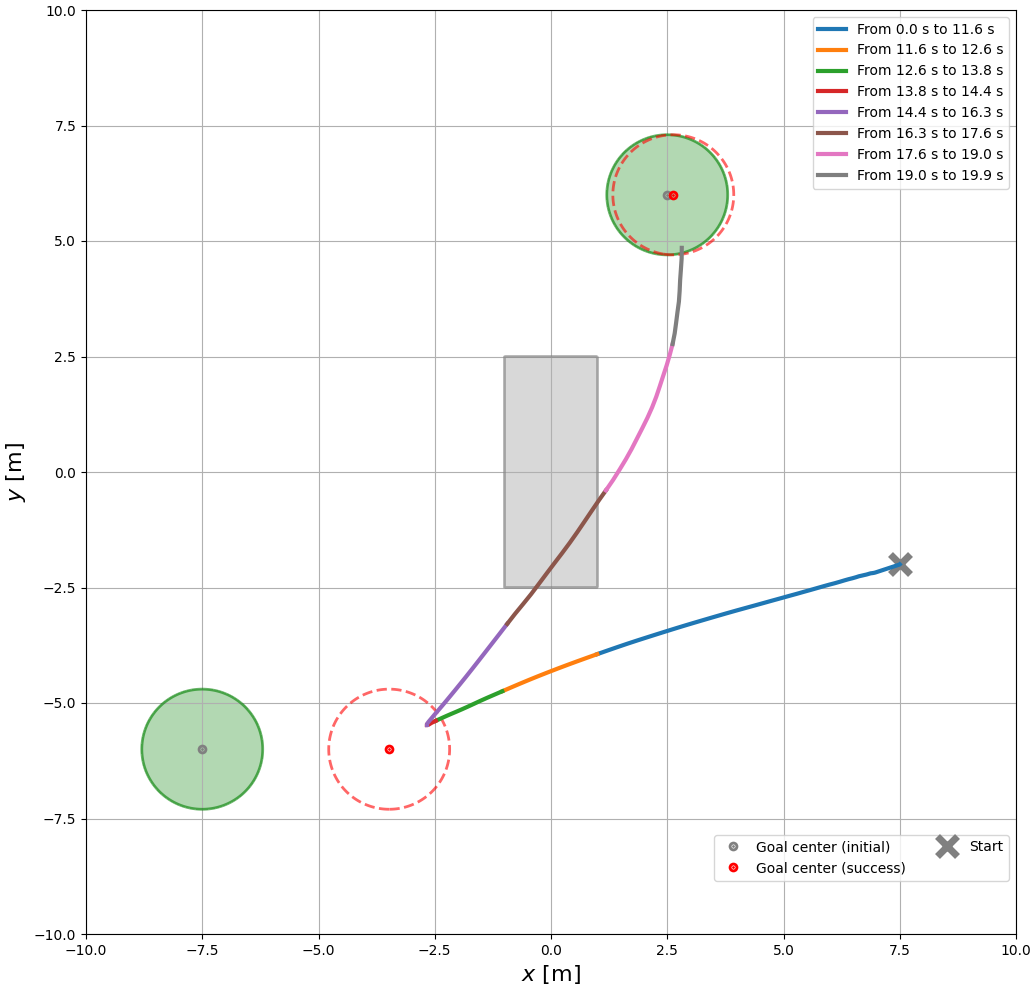}
    \end{minipage}
	}\subfigure[Spacecraft Rendezvous.] 
	{\label{fig:case3}
 \begin{minipage}[b]{0.33\linewidth}
	\centering
 \includegraphics[height=5.8cm]{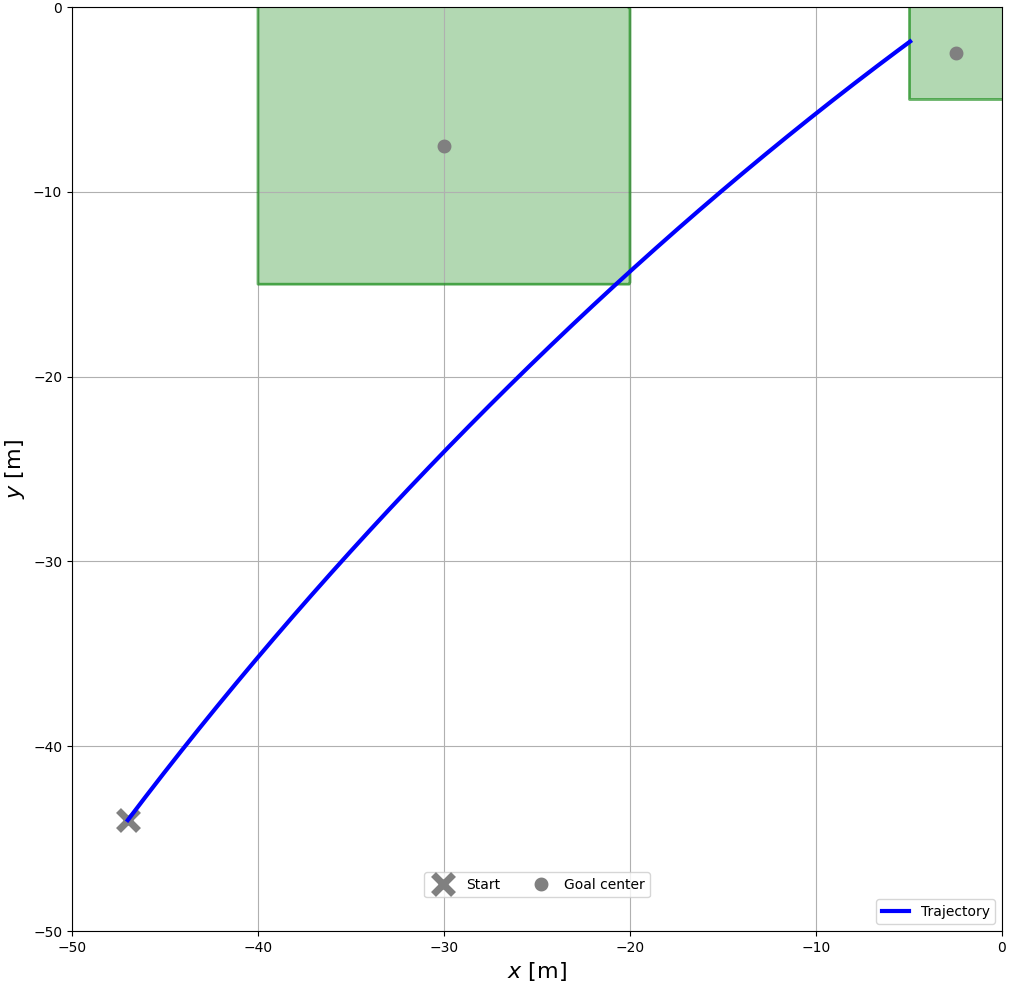}
    \end{minipage}
	}\subfigure[Unicycles Robots.] 
	{\label{fig:case4}
 \begin{minipage}[b]{0.33\linewidth}
	\centering
 \includegraphics[height=5.8cm]{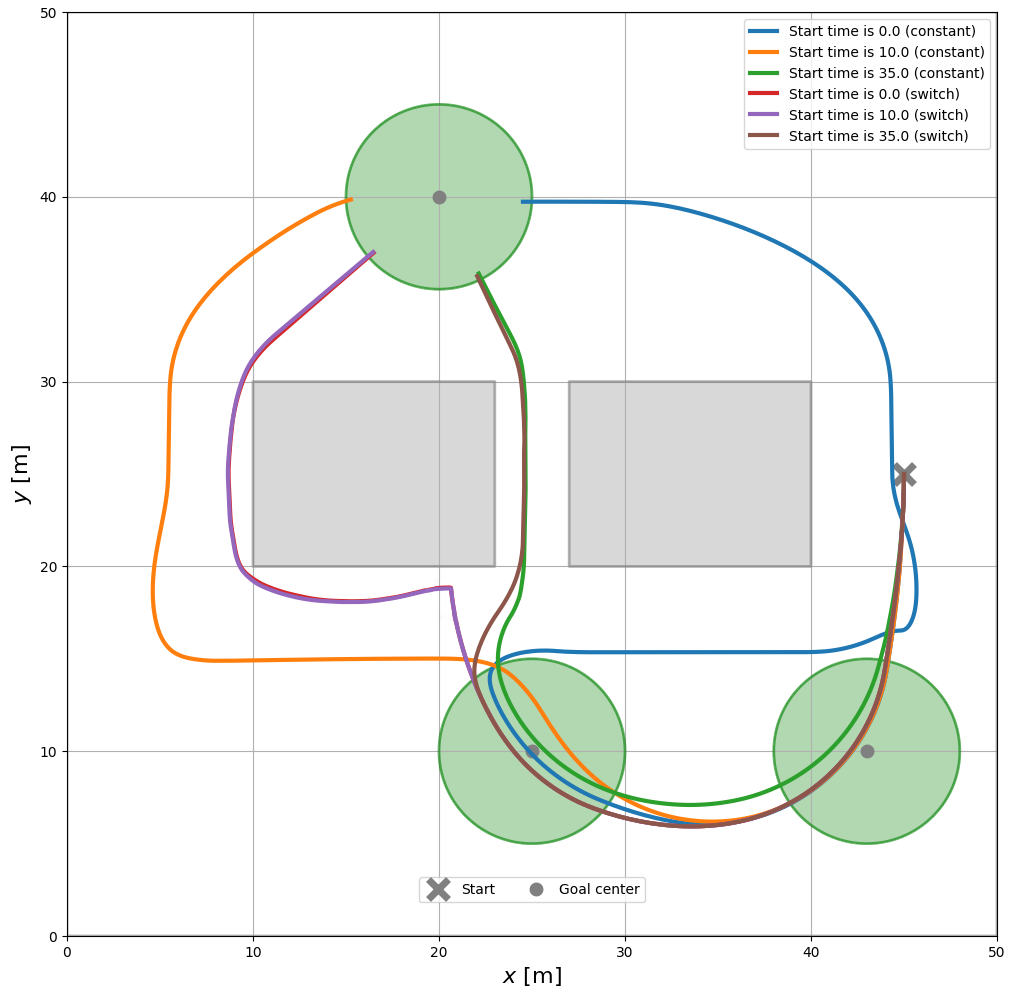}
    \end{minipage}
	}
    \caption{Simulation Results of Case Studies.}
   \label{fig:case2-4}
\end{figure*}

\subsection{Double Integrator}
Here we consider system with time varying dynamic. Specifically, state $x=[x, v_x, y, v_y]^\top$ denotes $x$-position, $x$-velocity, $y$-position, $y$-velocity and $u=[u_x, u_y]^\top$ denotes $x$-acceleration, $y$-acceleration, respectively. The system dynamic  is 
$\dot{x}=f(x)+g(x)u$, where
\begin{equation} \label{eq:doublesystem}
f(x)  =   \begin{bmatrix}
    v_x & 0 & v_y & 0\\
\end{bmatrix}^\top
, g(x)=
 \begin{bmatrix}
    0 & 0 & 0 & 1\\
    0 & 1 & 0 & 0
 \end{bmatrix}^\top. 
\end{equation}
 The control input set $\mathcal{U}(t)$ is a box over each dimension with time-varying parameter $u_{m}(t)$ defined as
 \begin{align*}
     \mathcal{U}(t)&=\{(u_x,u_y) \in \mathbb{R}^2 \mid |u_x| \leq u_{m}(t), |u_y| \leq u_{m}(t)  \},\\ u_{m}(t)& =
    \begin{cases}
       0.5+ 0.05t & 0 \leq t < 10 \\
        1 & t \geq 10 
    \end{cases}.
 \end{align*}
Such control input set may happen since the actuator experiences saturation during initial period.
The system can be converted to time-varying dynamic by modifying control input set as a constant set $\mathcal{U}=\{(u_x,u_y) \in \mathbb{R}^2 \mid |u_x| \leq 1, |u_y| \leq 1  \}$ and $g(x)$ in \eqref{eq:doublesystem} as
\begin{align*}
    g(x,t) =  \begin{bmatrix}
    0 & 0 & 0 & u_{m}(t)\\
    0 & u_{m}(t) & 0 & 0
 \end{bmatrix}^\top.
\end{align*}
The workspace of robot is a smart factory, where robot needs to get items from two circle dynamic agents. The dynamic agents will move along $x$-axis with period $10$s. There is another rectangle dynamic agent moving along $y$-axis with period $20$s and robot must avoid collision with it. The motion of agents along $x$-axis and $y$-axis, denoted by $\hat{x}$ and $\hat{y}$ respectively, satisfy that 
\begin{align*}
    \hat{x}(t) & =
    \begin{cases}
       t & 0 \leq t < 5 \\
        10-t & 5 \leq t \geq 10 \\
        \hat{x}(t-10) & t > 10
    \end{cases}, \\
    \hat{y}(t) & = \begin{cases}
        -1.5t & -5 \leq t <5 \\
        -15+1.5t & 5 \leq t \leq 15 \\
        \hat{y}(t-20) & t > 15
    \end{cases}.
\end{align*}
The initial position of target agents are $c_1 = (-7.5, -6)$ and $c_2 = (2.5,6)$. Dynamic target sets are defined by $R_1(t)=\{ (x,x_v,y,v_y) \in \mathbb{R}^4 \mid (x-c_1(1)-\hat{x}(t))^2+ (y-c_1(2))^2 \leq 1.3^2 \}$ and $R_2(t)=\{ (x,x_v,y,v_y) \in \mathbb{R}^4 \mid (x-c_2(1)-\hat{x}(t))^2+ (y-c_2(2))^2 \leq 1.3^2 \}$, respectively. Moreover, the rectangle dynamic obstacle, characterized by the lower left point $p_1^t=(-1,-2.5+\hat{y}(t))$ and upper right point $p_2^t=(1,2.5+\hat{y}(t))$ at time $t$, is defined by $G_o(t)=\{ (x,x_v,y,y_v) \in \mathbb{R}^4 \mid p_1^t(1) \leq x \leq p_2^t(1), p_1^t(2) \leq y \leq p_2^t(2) \}$.
The physically feasible state space of robot is $G_s=\{ (x,x_v,y,v_y) \in \mathbb{R}^4 \mid |x| \leq 10, |v_x| \leq 3, |y| \leq 10, |v_y| \leq 3 \}$.
Let $G(t)=G_s \setminus G_o(t)$. The state sets above at $t=0$ are illustrated in Figure~\ref{fig:case2}. 
The overall MRA task is described by
\[
\Phi=(0,20,\mathbb{T}=(R_1,R_2),\mathbb{G}=(G,G)).
\]
We convert each dynamic target and safe regions to value function using signed distance function. 
The initial state of robot is $(-7.5,0,-2,0)$ and the robot trajectory is shown in Figure~\ref{fig:case2}. The robot reach target $R_1$ and $R_2$ at $t_1=14.02$ and $t_2=19.87$, respectively. The positions of dynamic target at $t_1$ and $t_2$ are illustrated by red dot circles. From Figure~\ref{fig:case2}, robot may have collision with dynamic obstacle at coffee color or orange trajectory. However, when robot is at these points, time is larger than 10s, which means that $y$ coordinate of dynamic obstacle is larger than $0$. Thus no collision will occur and robot achieves the MRA task.

\subsection{Spacecraft Rendezvous}
We consider a spacecraft rendezvous example adopted from \cite{chan2017verifying,yu2024model}, where the Hill's relative coordinate frame, centered on the target spacecraft, is used to describe the planar motion of the chaser spacecraft on an orbital plane towards the target spacecraft. The dynamics of the system are governed by the following nonlinear equations:
\begin{align*}
            \dot{x} & = v_x \\
        \dot{y} & = v_y \\
        \dot{v}_x & = n^2 x + 2nv_y + \frac{\mu}{r^2} - \frac{\mu}{r_c^3}(r+x) +\frac{u_x+d}{m_c} \\ 
        \dot{v}_y & = n^2 y - 2nv_x-\frac{\mu}{r_c^3}y+ \frac{u_y+d}{m_c}
\end{align*}
where the state of the system is $x=[x,y,v_x,v_y]^\top$,
the   control input corresponds to the chaser's thrusters, denoted as
  $u=[u_x, u_y]^\top$, with each direction's maximum thrust limited to $10$N, i.e., the thruster forces are constrained as $u \in \mathcal{U} = [-10, 10] \times [-10, 10]$.
 Additionally, the disturbance is represented by $d \in \mathbb{R}$, which accounts for a 0.2\% error in the thruster force, i.e., $d \in \mathcal{D} = [-0.02, 0.02]$.
Other system parameters are given by: $\mu=3.986\times 10^{14}\times 30^2 [\text{m}^3/\text{min}^2]$, $r=42164\times10^3[\text{m}]$, $m_c=500[\text{kg}]$, $n=\sqrt{\frac{\mu}{r^3}}$ and $r_c=\sqrt{(r+x)^2+y^2}$. 

Assume the distance between the chaser and the target spacecraft is currently less than 50 m in each dimension.
The chaser should further approach to target spacecraft and maintain a low velocity.
Before docking to target spacecraft, chaser need to first reach specific region to further confirm the situation of the docking site by camera. 
Therefore, the   feasible state set is given by 
\[
G=\{ (x,y,v_x,v_y) \in \mathbb{R}^4 \mid x,y \in [-50,0], v_x,v_y \in [0,3] \}
\]
and the two target sets are defined by 
\begin{align}
    R_1 = \{ (x,y,v_x,v_y) \in \mathbb{R}^4 \mid x \in [-40,-20], y \in [-15, 0] \},\nonumber\\
    R_2 = \{ (x,y,v_x,v_y) \in \mathbb{R}^4 \mid x,y \in [-5,0], v_x,v_y \in [0,2] \}.\nonumber
\end{align}
 The overall MRA task is described by
\[
\Phi=(0,60,(R_1,R_2),(G,G)).
\]
We assume that the initial state is $(-47,-44,1.35,1.8)$. 
The simulation trajectory under disturbance is shown in Figure~\ref{fig:case3}, where the final velocity of the chaser is  $(v_x,v_y)=(1.77,1.31)$. 
Thus the MRA task is satisfied.

\subsection{Kinematic Unicycles Robots}
In this case study, we consider a mobile robot modeled by kinematic unicycles dynamic. 
Specifically, the state $[x,y,\theta]^\top$ denotes $x$-position, $y$-position and angle, respectively. 
The control input $[v,\omega]^\top \in \mathcal{U}$ denotes speed and angular velocity, respectively. 
The dynamic equation is given by
\begin{equation}
\dot{x} = v \mathsf{ cos } \theta, \quad \dot{y} = v \mathsf{ sin } \theta, \quad \dot{\theta} = \omega \nonumber
\end{equation}
such that $(v,\omega)\in\mathcal{U}=[0,3]\times [-0.3, 0.3]$.

We consider a scenario where the robot operates in a workspace with three circular target regions and two rectangular obstacles. 
The target regions are defined by centers at $(20, 40)$, $(25, 10)$, and $(43, 10)$, each with a radius of 5, denoted as $R_1$, $R_2$, and $R_3$, respectively. 
The rectangular obstacles, $G_1$ and $G_2$, are defined by their lower-left and upper-right corners: $p_{11} = (10, 20)$, $p_{12} = (23, 30)$ for $G_1$, and $p_{21} = (27, 20)$, $p_{22} = (40, 30)$ for $G_2$. These targets and obstacles are illustrated in Figure~\ref{fig:case4}. 
The robot operates in the workspace $G_f = [0, 50] \times [0, 50] \times [0, 2\pi]$, and the collision-free state space is defined as $G = G_f \setminus (G_1 \cup G_2)$.

The robot is required to visit the target regions in the order $R_3$, $R_2$, and $R_1$ within 60 seconds, while avoiding collisions. This task can be expressed as the following MRA task:
\[
\Phi=(0,60,(R_3,R_2,R_1),(G,G,G)).
\]
While the robot must complete the  task by the final time, we aim to minimize the time to complete it. 
For time-invariant systems and regions of interest, we modify the value function $\mathtt{b}$ from Algorithm~\ref{alg:procedure} by defining a translated version $\mathtt{b}'$ such that $\mathtt{b}'(x, t) = \mathtt{b}(x, t + t_0)$ for some $t_0 \geq 0$. 
This translation effectively reduces the latest arrival time by $t_0$. In our case study, we evaluate three scenarios with $t_0 = 0$, $t_0 = 10$, and $t_0 = 35$.

Additionally, we implement the control input set defined in \eqref{eq:modifiedfeasiblecontrolset} with $\beta = 1.2$. 
The control input sets in \eqref{eq:feasiblecontrolset} and \eqref{eq:modifiedfeasiblecontrolset} are denoted as ``constant" and ``switch“”, respectively, in Figure~\ref{fig:case4}.
We also consider a reference controller $(v_{\text{ref}}, \theta_{\text{ref}})$, where:\vspace{-6pt}
\begin{itemize}
\item $v_{\text{ref}} = k_v d_o(x)$ scales linearly with the distance $d_o(x)$ to obstacles,\medskip
\item $\theta_{\text{ref}} = k_\theta \theta_e(x)$ adjusts based on the angle $\theta_e(x)$ to the current target center.
\end{itemize}
Here, both $k_v$ and $k_\theta$ are constant gains.

The   initial state of the robot is $x_0 = (45, 25, 1.5\pi)$, and its trajectories under different control laws and parameters are shown in Figure~\ref{fig:case4}. 
When comparing different values of $t_0$ under the same control input set, the robot chooses a shorter path with higher $t_0$. 
In contrast, when comparing different control input sets under the same $t_0$, using the control input set defined in \eqref{eq:modifiedfeasiblecontrolset}, the robot stays closer to the obstacle. This is because, when the distance between the obstacles and the robot exceeds the pre-defined value $\beta = 1.2$, the modified control input set allows the robot to approach obstacles if the reference controller suggests doing so.

\section{Conclusion} \label{sec:con}
In this paper, we addressed the problem of synthesizing controller for multiple reach-avoid tasks in nonlinear time-varying systems subject to disturbances. We demonstrated how the feasibility of these tasks can be verified through a series of value functions computed using the Hamilton-Jacobi reachability  method.
Furthermore, we proposed an online procedure to utilize these value functions for achieving the multiple reach-avoid tasks.
Additionally, we explored how the techniques developed for the MRA task can be leveraged to solve the controller synthesis problem for linear temporal logic tasks.
Extensive experiments were conducted to demonstrate the effectiveness of the proposed method.
In the future, we aim to extend our work to the synthesis of multiple reach-avoid tasks with probabilistic guarantees for stochastic systems. We also plan to employ learning-based methods to address the scalability challenges in solving HJR  PDEs for high-dimensional systems.

\bibliographystyle{plain}
\bibliography{main} 
\end{document}